# Formation of crystalline bulk heterojunctions in organic solar cells: insights from phase-field simulations

*Olivier J.J. Ronsin\*[a,b] and Jens Harting[a,b,c]*

a. Helmholtz Institute Erlangen-Nürnberg for Renewable Energy, Forschungszentrum Jülich, Fürther Straße 248, 90429 Nürnberg, Germany

b. Department of Chemical and Biological Engineering, Friedrich-Alexander-Universität Erlangen-Nürnberg, Fürther Straße 248, 90429 Nürnberg, Germany

c. Department of Physics, Friedrich-Alexander-Universität Erlangen-Nürnberg, Fürther Straße 248, 90429 Nürnberg, Germany

## Abstract

The performance of organic solar cells strongly depends on the bulk-heterojunction (BHJ) morphology of the photoactive layer. This BHJ forms during the drying of the wet-deposited solution, because of physical processes such as crystallization and/or liquid-liquid phase separation (LLPS). However, the process-structure relationship remains insufficiently understood. In this work, a recently developed, coupled phase-field –fluid mechanics framework is used to simulate the BHJ formation upon drying. For the first time, this allows to investigate the interplay between all the relevant physical processes (evaporation, crystal nucleation and growth, liquid demixing, composition-dependent kinetic properties), within a single coherent theoretical framework. Simulations for the model system P3HT-PCBM are presented. The comparison with previously reported in-situ characterization of the drying structure is very convincing: the morphology formation pathways, crystallization kinetics, and final morphology are in line with experimental results. The final BHJ morphology is a subtle mixture of pure crystalline donor and acceptor phases, pure and mixed amorphous domains, which depends on the process parameters and material properties. The expected benefit of such an approach is to identify physical design rules for ink formulation and processing conditions to optimize the cell's performance. It could be applied to recent organic material systems in the future.

# Introduction

**Context.** Organic solar cells are a promising photovoltaics technology. They can be produced using solution processing, a cheap, scalable and ecologically-friendly method. This also allows for the fabrication of flexible or semi-transparent modules, opening the way to new applications like building-integrated or portable photovoltaics. However, the efficiency of such solar cells needs to be improved. In parallel to the development of new photoactive materials, understanding and improving the processing route is promising, because the nanomorphology of photoactive layers required for high device performance forms during the fabrication process. In organic solar cells (OSC), the photoactive layers are typically $100 - 300\ nm$ thick and made of two materials, one electron donor (frequently a polymer) and one electron acceptor. The current understanding of the structure-property relationship can be summarized as follows: [1] [2] [3] [4] light absorption leads to the formation of strongly bound excitons, which need to be separated into free charge carriers. This occurs usually in a mixed interphase between a donor-rich region and an acceptor-rich region. Since the excitons' mean free path is about $10\ nm$, the domain size of the donor and acceptor regions should not exceed $10 - 20\ nm$. After the separation, the electrons and holes have to be efficiently transported to the electrodes. Thus, charge transport should occur in relatively pure and crystalline donor and acceptor phases, respectively. In addition, the charge carriers should be provided pathways to their respective collecting electrode, either through the pure phases or a mixed phase, provided its composition allows for charge transport. Therefore, the desired structure is a so-called 'bulk heterojunction' (BHJ), a co-continuous nanostructure of pure, significantly crystalline donor and acceptor phases eventually bridged with mixed phases.

The bulk heterojunction concept has led to very successful results over the last 20 years, with the best organic solar cell efficiencies now reaching $16 - 18\%$ [5] [6] [7]. These results have been obtained by developing new donor and acceptor materials (notably non-fullerene acceptors [4] [8] [9]) with improved optoelectronic properties, but also by tuning their thermodynamic properties and the processing conditions. This is because the BHJ formation is controlled by thermodynamic processes (crystallization, liquid-liquid phase separation (LLPS)) occurring during the fabrication of solution-processed OSC.[10] [11] [12] [13] [14] [15] [16] These phenomena are triggered by the concentration increase upon drying of the wet film. At the end of the drying, the morphology evolution is kinetically quenched far from its thermodynamic equilibrium due to vanishing solvent concentration. However, the process-structure relationship remains poorly understood: first, the assessment of the film morphology at the nanoscale, with many different phases, in blends of purely organic materials, is an experimental challenge. [17] [18] [19] Second, new materials are frequently developed for OSC applications so that a large number of different blends is investigated, which might hinder the determination of a coherent picture.[9] Third, the BHJ formation strongly depends on the thermodynamic and kinetic properties of the donor-acceptor-solvent(s) mixture, which makes it a very complicated coupled thermodynamic and kinetic problem and thus a theoretical challenge.[19] [20] [21]

**Objectives.** The present paper is a contribution to the progress on the theoretical understanding of BHJ formation. The idea is to use a unified simulation framework handling the minimal physics required to describe the problem. First, crystal nucleation and growth of donor and acceptor materials, LLPS, and evaporation of solvents are the basic phase transformations to be taken into account. Second, the film composition is constantly changing due to solvent removal, and the kinetic properties defining the evolution rate of the morphology (diffusion coefficients for diffusive mass transfer, viscosities for advective mass transfer, crystallization rates) might vary over orders of magnitude. Their composition-dependence has to be integrated in the model. Third, the typical length scales to be described range from a couple of nanometres to more than one micrometre for the wet film thickness and typical processing times are, for example, at least in the range of seconds or minutes for the drying step. Fourth, the expected presence of many different liquid and solid domains with various compositions requires a straightforward handling of all interfaces between the phases.

Even if some very interesting insights have been obtained by simulations at small scales using molecular dynamics or dissipative particle dynamics,[22] [23] [24] [25] [26] the length and times scales of the problem, as well as the acknowledged importance of kinetics makes continuum mechanics methods more appropriate to tackle the problem. Among these, phase-field (PF) simulations are particularly well-appropriated, because they are basically thought to describe the kinetic evolution of a multiphase system towards its thermodynamic equilibrium, using a diffuse interface approach.[27][28][29][30][31] PF is a well-established and versatile method which has been used to model LLPS, crystallization in various systems, precipitation, liquid-vapour transformations and many more, and which has been coupled to fluid mechanics (FM) equations. [32][33][34][35][36][37][38][39][40][41][42] In the fields of OSC, it has already been applied to investigate the evolution of a dry amorphous donor-acceptor mixture.[43][44] A significant step towards understanding of the morphology formation upon drying has been done by simulating evaporation-induced spinodal decomposition.[45][46][47][48][49][50][51] Unfortunately, the crystallization was not taken into account in these works. However, in another research field (drug delivery applications), Saylor and Kim simulated successfully simultaneous LLPS and crystallization in a drying ternary polymer mixture, [52][53][54][55] even if several processes like crystal impingement or advection were not taken into account. Building on these seminal works, we recently proposed a new PF framework taking into account all the physical features listed above. [56][57][58]

In the present paper, we apply this PF model to simulate the BHJ formation upon drying of a polymer-small molecule organic photoactive layer. A major improvement as compared to previous studies is that crystallization is taken into account: of course, the morphology formation mechanisms and the final structure are significantly affected. The objectives are to test the ability of our framework to handle "realistic" OSC systems with appropriate material parameters, to understand the BHJ formation pathways, to quantitatively characterize the morphology evolution, and to compare with experimental results for validation.

**Relevant questions for the morphology formation pathways.** In general, depending on the material properties, several phase transformations can take place in the drying ternary donor-acceptor-solvent solution: liquid-liquid demixing due to donor-acceptor, donor-solvent or acceptor-solvent immiscibility, donor crystallization, and acceptor crystallization. Since these processes are not instantaneous and since the mobilities in the drying mixture decreases tremendously upon drying, the time provided for significant phase transformation is limited. Therefore, identifying qualitatively the morphology formation pathway requires first of all to find out whether these processes are thermodynamically expected, if they have sufficient time to occur and at which solvent concentration this would be.

Furthermore, one needs to understand in which order and at which vertical position, from the substrate to the film surface, they appear. Of course, interactions between the different phase transformations, for instance LLPS-induced crystallization or conversely crystallization-induced LLPS can complicate the morphology formation. As will be illustrated below, the analysis of the phase diagrams and of the characteristic time scales for the various rate processes (evaporation, crystal nucleation, crystal growth, grain coarsening, diffusion, advection, spinodal decomposition, liquid phase coarsening,) provides very useful hints, even if one has to keep in mind that these time scales are actually composition- and thus time-dependent. Questioning the characteristic length scales (initial wet film height, dry film height, critical radius of emerging nuclei, minimal size of demixed regions for spinodal decomposition) also helps to understand how the phase transformations take place. For instance, if the minimal size for liquid-liquid demixing is comparable with the film height, this might result in vertical layering or even hinder the LLPS in the z-direction,[33] and vertical concentration gradients due to diffusional limitations are more likely in thick films. Indeed, the vertical organization of the photoactive layer is a sophisticated problem, which is determined by all the aspects described above. It will certainly be a long-standing research topic to sort out the various relevant morphology formation pathways for OSC photoactive layers, and this cannot be the topic of the current paper. Instead, we focus in the following on one particular case, basing on the well-known P3HT-PCBM system.

## Model

### Kinetic equations

**Field variables.** Our coupled Phase Field – Fluid Mechanics (PF-FM) model has been presented extensively in a previous paper the reader is referred to for more details.[58] Here, we describe only the main features and underlying equations of the simulation method.

The system of $n$ materials (here: donor, acceptor, one solvent and the air), among which $n_c$ can crystallize (here: donor and acceptor) is described by the $n$ volume fraction fields $\varphi_i$, the $n_c$ order parameter fields $\phi_k$ varying between 0 and 1 from the amorphous/liquid to the crystalline/solid phase, $n_c$ marker fields $\theta_k$ for identification of single crystallites. ($\theta_k$ being undefined outside the crystals), one order parameter field $\phi_v$ varying between 0 and 1 from the amorphous to the vapour phase, and a single velocity field $\boldsymbol{v}$. The density is assumed to be constant and homogeneous in the whole simulation box.[57][58]

**Mass transport.** The evolution equation for the volume fractions is given by the advective Cahn-Hilliard-Cook equation [59][60][61] for $i = 1 \dots n-1$:

$$\frac{\partial \varphi_i}{\partial t} + \boldsymbol{v}\boldsymbol{\nabla}\varphi_i = \frac{v_0}{RT}\nabla\left[\sum_{j=1}^{n-1}\Lambda_{ij}\boldsymbol{\nabla}\left(\frac{\delta G}{\delta\varphi_j} - \frac{\delta G}{\delta\varphi_n}\right)\right] + \zeta_{CH}{}^i \quad (1)$$

and the last volume fraction is deduced from the required volume conservation, $\sum \varphi_i = 1$. This is a generalized, multicomponent advection-diffusion equation where the exchange chemical potentials act as driving forces. The potentials are obtained from a free energy functional $G$ including surface tension contributions which will be detailed below, $\delta G/\delta \varphi_j$ standing for the functional derivative of the free energy. $R$ is the gas constant, $T$ the temperature and $v_0$ the molar volume of the lattice site as defined in the Flory-Huggins theory.[62][63] $\Lambda_{ij}$ are field-dependent symmetric Onsager mobility coefficients which are related to the self-diffusion coefficients of each material by the slow-mode theory.[64] The composition-dependence of the self-diffusion coefficients in the liquid phase will be discussed below. $\zeta_{CH}{}^i$ are coupled Gaussian noise terms whose expression can be found in[58]

**Evaporation.** The evolution equation for the vapour order parameter $\phi_v$ is given by the advective Allen-Cahn equation

$$\frac{\partial \phi_v}{\partial t} + \boldsymbol{v}\boldsymbol{\nabla}\phi_v = -\frac{v_0}{RT}M_v\frac{\delta G}{\delta \phi_v} \quad (2)$$

where $M_v$ is the Allen-Cahn mobility coefficient for the vapor. The Allen-Cahn equation is non-conservative so that during the simulation of a drying film, the condensed phase region shrinks upon solvent liquid-vapour phase transition. The evaporation flux of each solvent is defined at the top boundary of the simulation domain by

$$j_{i,HK} = \alpha\sqrt{\frac{v_0}{2\pi RT}\frac{N_i}{\rho_i}}P_0(\varphi_i^{vap} - \varphi_i^\infty) \quad (3)$$

where $\varphi_i^\infty$ is defined as $\varphi_i^\infty = P_i^\infty/P_0$ with $P_i^\infty$ being the partial pressure in the environment, $P_0$ is a reference pressure, $\varphi_i^{vap}$ is the volume fraction of material $i$ in the vapor phase close to the film, and $\alpha$ is the evaporation-condensation coefficient. The solvent leaving the simulation box is replaced by a buffer material which remains in the gas phase only. Using Equations 2 and 3 with a high mobility coefficient $M_v$, evaporation is a two-step process with a fast LV transition for the solvents, the build-up of local LV equilibrium at the film surface, and a slow diffusion process away from the equilibrium vapour layer. The evaporation kinetics is fully driven by the diffusive flux (Equation 33) so that the Hertz-Knudsen theory of evaporation [65] is recovered.[57]

**Crystallization.** The evolution of the order parameters $\phi_k$ for the $n_c$ crystalline materials is given by the stochastic Allen-Cahn equations

$$\frac{\partial \phi_k}{\partial t} + \boldsymbol{v}\boldsymbol{\nabla}\phi_k = -\frac{v_0}{RT}M_k\frac{\delta G}{\delta \phi_k} + \zeta_{AC}{}^k \quad (4)$$

Here, $\zeta_{AC}{}^k$ are Gaussian noise terms which allow for spontaneous nucleation events. The Allen-Cahn mobility coefficient for crystals is $M_k = M_{k,0}\,D_{s,k}^{liq}(\{\varphi\})/D_{s,kk}^{liq}$ (where $D_{s,kk}^{liq}$ is the self-diffusion coefficient of material $k$ in a matrix made of pure $k$, $M_{k,0}$ the Allen-Cahn mobility in the pure material $k$ and $D_{s,k}^{liq}(\{\varphi_k\})$ the self diffusion coefficient in the mixture) so that the crystallization process is favoured in highly mobile media like dilute solutions, as will be discussed below. In addition to this, a marker value $\theta_k$ is attributed randomly to each newly created nucleus as soon as the order parameter $\phi_k$ and the volume fraction $\varphi_k$ both exceed given thresholds. Upon crystal growth, the newly crystallizing nodes at crystal boundaries are given the marker value of the crystal core. Finally, as soon as the volume fraction or order parameter drop below the threshold, the marker value becomes undefined again. With such a simple procedure, the marker value is uniform in a given crystal and is bound to the evolution of the order parameter and volume fraction field.

**Fluid flows.** The Allen-Cahn and Cahn-Hilliard equations lead to the minimization of the free energy, so that the kinetic evolution of system towards its thermodynamic equilibrium is calculated. In order to take into account the role of advective mass transport, they are coupled to fluid mechanics equations. In such a small system, the Reynolds number is always small and fluid inertia can be neglected. In addition, buoyancy forces are negligible as compared to the capillary forces generated at the many interfaces in the film. Assuming also incompressibility, the continuity and momentum conservation equations read

$$\begin{cases}\nabla \boldsymbol{v} = 0 \\ -\boldsymbol{\nabla}P + \boldsymbol{\nabla}(2\eta_{mix}\boldsymbol{S}) + \boldsymbol{F}_\varphi + \boldsymbol{F}_\phi = \boldsymbol{0}\end{cases} \quad (5)$$

Here, $P$ is the pressure, $\mathbf{S}$ is the strain rate tensor and $\eta_{mix}$ a composition-dependent viscosity given by $\eta_{mix} = \prod_{i=1}^{n} \eta_i^{\varphi_i}$ in the liquid, $\eta_i$ being the viscosities of the pure materials. The viscosity in the crystals (resp. vapor phase) are set to be significantly higher (resp. lower) than in the liquid. $\mathbf{F}_\varphi$ and $\mathbf{F}_\phi$ are the capillary forces arising at interfaces from the volume fraction and order parameter gradients, respectively:[41][58][66]

$$\begin{cases} \mathbf{F}_\varphi = \nabla \left[ \sum_{i=1}^{n} \kappa_i (|\nabla \varphi_i|^2 \mathbf{I} - \nabla \varphi_i \times \nabla \varphi_i) \right] \\ \mathbf{F}_\phi = \nabla \left[ \sum_{k=1}^{n_c} \varepsilon_k^2 (|\nabla \phi_k|^2 \mathbf{I} - \nabla \phi_k \times \nabla \phi_k) \right. \\ \left. \qquad + \varepsilon_v^2 (|\nabla \phi_v|^2 \mathbf{I} - \nabla \phi_v \times \nabla \phi_v) \right] \end{cases} \quad (6)$$

Here $\kappa_i$, $\varepsilon_k$ and $\varepsilon_v$ are surface tension parameters for liquid-liquid (LL), liquid-solid (LS) and liquid-vapour (LV) interfaces which fix the interfacial tensions and the intensity of the capillary forces.

**Thermodynamics: the free energy**

The thermodynamic behaviour is encoded in the free energy functional $G$ which contains different contributions:

$$G = \int_V dV \begin{bmatrix} (1 - p(\phi_v)) \left( G^l + \sum_{k=1}^{n_c} G_k^c \right) + p(\phi_v) G^v + \sum_{k=1}^{n_c} G_k^{cv} \\ + G^n + G^i \end{bmatrix} \quad (7)$$

$G^l$ is the LL free energy density of mixing in the condensed phase, $G_k^c$ is the contribution related to the LS phase transition of material $k$, $G^v$ is the free energy density in the gas phase, $G_k^{cv}$ are crystal-vapour interaction terms, $G^n$ is a purely numerical contribution and $G^i$ the interfacial tension energy density. At the LV interface, the free energy is interpolated between the condensed phase and the vapour phase value using the classical phase-field interpolation function $p(\phi) = \phi^2(3 - 2\phi)$. For the LL free energy density of mixing, we use the Flory-Huggins theory:[62][63]

$$G^l = \frac{RT}{v_0} \left( \sum_{i=1}^{n} \frac{\varphi_i \ln \varphi_i}{N_i} + \sum_{i=1}^{n} \sum_{j>i}^{n} \varphi_i \varphi_j \chi_{ij,ll} \right) \quad (8)$$

Here, $N_i$ represents the molar volume $v_0 N_i$ of each material and $\chi_{ij,ll}$ are the binary Flory-Huggins interaction parameters in the liquid state. Following previous works, [30][67][68] the free energy density of crystallization is given by

$$G_k^c = \begin{array}{l} \rho_k \varphi_k^2 (g(\phi_k) W_k + p(\phi_k) L_k^c) \\ + \dfrac{RT}{v_0} \sum_{j \neq k}^{n} \phi_k^2 \varphi_k \varphi_j \chi_{kj,sl} \end{array} \quad (9)$$

In the equation above, the first line of the RHS stand for the energy change upon crystallization. $\rho_k$ is the material density, $W_k$ fixes the height of the energy barrier upon phase change together with the double-well function $g(\phi) = \phi^2(\phi - 1)^2$, and $L_k^c$ is the heat of crystallization calculated from the heat of fusion $L_k^{fus}$ and the melting temperature $T_{m,k}$ using $L_k^c = L_k^{fus}(T/T_{m,k} - 1)$. The second line stands for modified pair interaction energies in the crystal between the crystallized molecules of material $k$ and amorphous molecules of material $j$. The solid-liquid Flory-Huggins interaction parameters $\chi_{kj,sl}$ fix the intensity of this augmented contributions. The vapour phase is assumed to be an ideal gas, so that defining $\varphi_{sat,i} = P_{sat,i}/P_0$ from the vapor pressures $P_{sat,i}$, the free energy density of the gas phase simply reads

$$G^v = \frac{RT}{v_0} \sum_{i=1}^{n} \varphi_i \ln \left( \frac{\varphi_i}{\varphi_{sat,i}} \right) \quad (10)$$

In addition, $G_k^{cv} = E_k \phi_k^2 \phi_v^2$ is a crystal-vapor interaction energy density whose intensity is given by the solid-vapour (SV) energy $E_k$. This term is active in the crystalline regions only (defined as nodes where the marker value is defined) and prevents the overlap of vapour with crystalline order parameters which might occur due to the diffuse interface nature of the PF method, and which could lead to unphysical crystal instability at the film surface. More details on this topic can be found in [58]. $G^n = \beta \sum_{i=1}^{n} \varphi_i^{-1}$ is a contribution preventing the volume fraction values to escape the physically meaningful ]0,1[ interval, and thus providing enhanced numerical stability and efficiency.[52] $\beta$ is chosen small enough to avoid any significant modification of the mixture physical properties. Finally, the interfacial, non-local energy density is related to the gradients of the field variables by

$$G^i = \begin{array}{l} \sum_{i=1}^{n} \dfrac{\kappa_i}{2} (\nabla \varphi_i)^2 + \dfrac{\varepsilon_v^2}{2} (\nabla \phi_v)^2 \\ + \sum_{i=1}^{n_c} \left( \dfrac{\varepsilon_k^2}{2} (\nabla \phi_k)^2 + p(\phi_k) \dfrac{\pi \varepsilon_{g,k}}{2} |\nabla| \delta(\nabla \theta_k) \right) \end{array} \quad (11)$$

Note that the $\nabla \varphi_i$ term will contribute to the interfacial tension of all kinds of interfaces, because there is a compositional variation within any interface. This is the only contribution for LL interfaces, but the LV interfacial energy also includes the $\nabla\phi_v$ term, the SL interfacial energy the $\nabla\phi_k$ term, and the SV interfaces both $\nabla\phi_v$ and $\nabla\phi_k$ terms. The $\nabla\theta_k$ contribution is a grain boundary energy term active at the interface between different crystals (i.e. having different marker values) of the same materials, $\varepsilon_{g,k}$ being the grain energy coefficient. This allows for a proper simulation of crystal impingement and grain coarsening.

# Simulations of photoactive film drying

## Thermodynamic and kinetic properties

In this section, we report on simulations of the morphology formation upon drying for a typical OSC photoactive layer. The target system is the P3HT-PC$_{61}$BM donor-acceptor blend dissolved in DCB, because it is a well-known system with a large amount of experimental data available, regarding the material parameters as well as the morphology characterization. Before turning to the simulations of film drying themselves, we review the thermodynamic and kinetic properties of the material system. The objective is to provide some simple guidelines and numbers that are of fundamental importance to understand the complex behaviour upon drying. The phase diagram of the ternary mixture, its drying kinetics and the crystallization kinetics of each solute in a binary, non-evaporating solution are discussed. Finally, we comment on the parameter set we choose for these simulations, and its relationship to the data available from the literature for the P3HT-PCBM-DCB mixture. All parameters are gathered in the Supporting Information S1.

**Thermodynamics: calculation of the phase diagrams.** The temperature-dependent binary donor-solvent, acceptor-solvent and donor-acceptor phase diagrams, as well as the ternary phase diagram at $60\ °C$ are shown in **Figure 1**. In order to determine this phase-diagram, the order parameter values minimizing the free energy density of the condensed phase $G^l + G_d^c + G_a^c$ at each blend composition are calculated for both donor and acceptor (subscripts $d$ and $a$, respectively). Then, the unstable region of the phase diagram and the equilibrium compositions are calculated using the derivative-free procedure proposed by Horst.[69] [70] The form chosen for the free energy allows to generate a multitude of different phase diagrams.[67] [68] [71]

The binary donor-solvent and acceptor-solvent phase diagrams are shown in **Figure 1a** and **Figure 1c** respectively, featuring a liquid-solid equilibrium. In addition to using parameters as close as possible to the ones measured for P3HT-DCB and PCBM-DCB (see discussion below), both phase diagrams are adapted considering two guidelines. First, the solubility limits have been measured to be $15\ mg \cdot mL^{-1}$ (resp. $42\ mg \cdot mL^{-1}$) at room temperature, [72] which corresponds to volume fractions of $1\%$ and $3\%$ for the liquids. They are known to increase with temperature, the PCBM solubility remaining higher until P3HT significantly increases around $50-60\ °C$ and becomes higher than the PCBM solubility. Even if slightly different results are reported elsewhere in the literature,[10] our phase diagrams aim at reproducing these trends qualitatively. Second, P3HT is known to crystallize upon drying almost as soon as the solubility limit is reached. [10] [17] Unfortunately, as will be illustrated below, in our PF simulations the solute volume fraction for the onset on nucleation $\varphi_d^{nucl,onset}$ (resp. $\varphi_a^{nucl,onset}$) is significantly higher than the liquidus volume fraction. It is related to the lowest volume fraction for which the mixture is unstable $\varphi_d^{min,unst}$ (resp. $\varphi_a^{min,unst}$). The polymer-solvent phase diagram is adjusted in order to have low values for $\varphi_d^{nucl,onset}$, and thus $\varphi_d^{min,unst}$. A side effect of this choice is that the solidus composition corresponds to crystals with significant amounts of solvent.

For the binary donor-acceptor blend (**Figure 1d**), the P3HT-PCBM eutectic phase diagram measured on the first DSC heating curve are recovered. [73] [74]At a typical processing temperature below $100\ °C$, a solid-solid equilibrium with nearly pure crystals of both materials is expected. In other words, the thermodynamically stable state of the P3HT-PCBM photoactive layer is a two phase, fully crystalline morphology of pure P3HT and PCBM crystals (remember that the semi-crystalline nature of the polymer is not taken into account in this work, and note that we talk about PCBM "crystals" even if the ordered nature of the fullerene aggregates is still open to debate).

Finally, the corresponding ternary donor-acceptor-solvent phase diagram at $60\ °C$ is shown in Figure 1b. The unstable domains arise due to the crystallization properties of the donor and the acceptor, hiding the liquid-liquid immiscibility region. Upon drying, a ternary liquid-solid-solid phase separation is expected between a crystalline, PCBM-free P3HT phase including small amount of solvents, a crystalline, P3HT-free PCBM phase including even less solvent and a liquid, P3HT-free, solvent phase including PCBM. Upon further drying, the situation switches to a binary solid-solid equilibrium with a P3HT crystalline phase including solvent but no PCBM, and a crystalline PCBM phase including less solvent. The amount of solvent in both phases goes down with diminishing overall solvent amount.

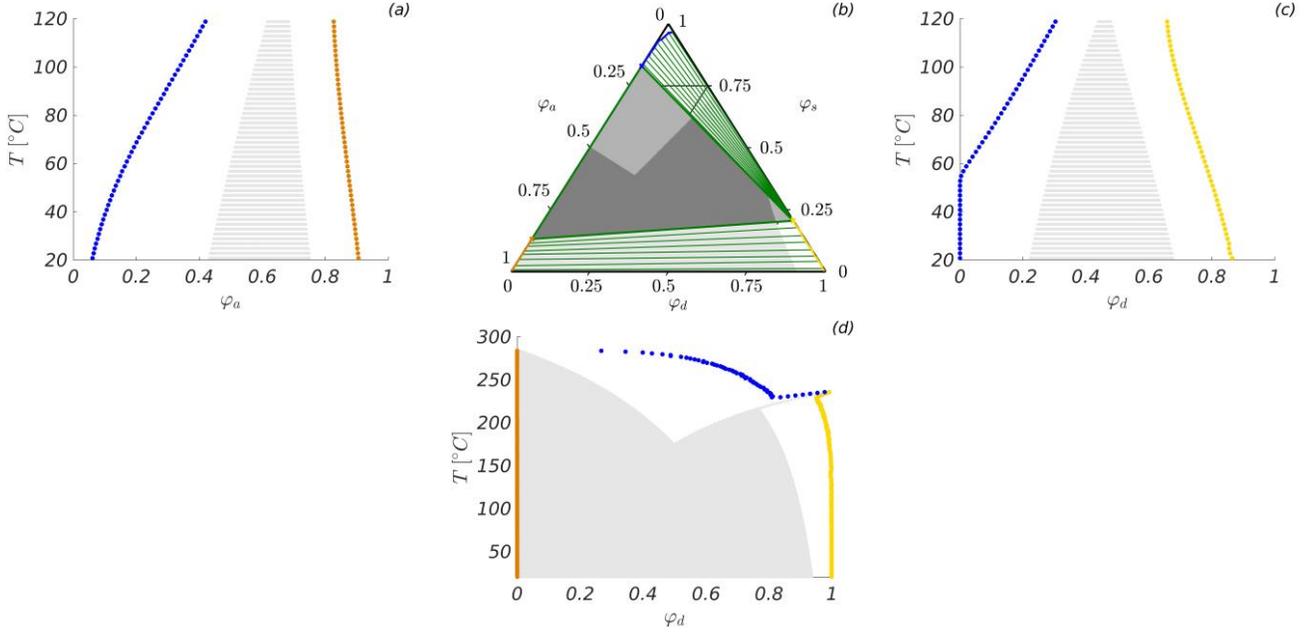

*Figure 1: phase diagrams of the investigated mixture (a) binary acceptor –solvent phase diagram (b) ternary donor-acceptor-solvent phase diagram at $60\,°C$ (c) binary donor –solvent phase diagram (d) binary donor –acceptor phase diagram. The unstable compositions are represented in grey, the liquidus in blue, the acceptor solidus in orange, the donor solidus in yellow and (for the ternary phase diagram) the tie lines in green. . The unstable compositions associated with a binary equilibrium are represented in light grey and (for the ternary phase diagram) the metastable compositions associated with a ternary equilibrium in middle grey, and the unstable compositions associated with a ternary equilibrium in dark grey. The material parameters used for the phase diagrams can be found in the Supporting Information S1, Table S1.*

**Kinetics: crystal growth and nucleation rates, evaporation.** Whether the morphology formation upon drying really follows the trends deduced from the equilibrium thermodynamics considerations detailed above depends on the kinetic properties of the mixture. To shed some light on this, we first simulate the crystallization process of each solution in a DCB solution at different, fixed volume fractions, starting from a homogenous crystal-free mixture (Figure 2). The nucleation rate $v_{nucl}$ is expected to be the product of three terms of a material *k* and can be written as the product of three terms:

$$v_{nucl} \propto M_k A(H_k) e^{-\frac{\Delta G^*}{RT}} \qquad (12)$$

Here, $H_k$ is the height of the energy barrier for the liquid-solid.[58] The last term is therefore a purely thermodynamic factor where $\Delta G^*$ is the energy barrier to be overcome for the formation of a stable nucleus. The second term is related to the probability of a fluctuation overcoming the energy barrier and depends only on the thermodynamic properties of the blend. Both terms contribute to a significant decrease of the nucleation rate with decreasing solute concentration. Finally, the first factor given by the Allen-Cahn mobility is a purely kinetic factor reflecting the local mobility of the molecules in the mixture. Remember that $M_k$ is assumed to be proportional to the self-diffusion coefficient in the mixture. In a solution, the mobilities strongly increase upon dilution. The balance with the thermodynamic contribution leads to a maximum nucleation rate at intermediate concentrations. The same holds for the crystal growth rate, the effect of increasing mobilities upon dilution being balanced by the decreasing thermodynamic driving force for phase change.

The composition-dependence of the diffusion coefficients is a very complex scientific question, especially for polymer solutions. In this work, however, we use for simplicity a logarithmic mean, $D_{s,i}^{liq}(\{\varphi\}) = \prod_{k=1}^{n}(D_{s,i}^{\varphi_k \to 1})^{\varphi_k}$, where $D_{s,i}^{\varphi_k \to 1}$ is the self-diffusion coefficient of the (liquid) $i^{th}$-material in the $k^{th}$ pure (liquid) materials. Setting the diffusion coefficients of the polymer donor (resp. the small-molecule acceptor) in the pure donor (resp. acceptor) roughly five (resp. three) decades below the self-diffusion coefficients at infinite dilution, the diffusion coefficients drop of about one decade every 20% (resp. 33%) volume fraction (see Supporting Information S1 for the exact parameter set). Using these values together with the thermodynamic properties presented above, we find that the crystallization rate of the polymer donor increases upon dilution due to the increasing mobility until $\varphi_d^{nucl,onset} \approx 0.2$, and then abruptly drops due to the thermodynamic properties (**Figure 2a**). In addition to the critical germ radius getting always larger and the driving force for crystallization always smaller with increasing dilution, emerging coupled order-parameter-volume fraction fluctuations tend to be smeared out instead of developing into germs, because they cannot reach the unstable region of the phase diagram. As a consequence, the nucleation rate drops over orders of magnitude for concentrations significantly lower than $\varphi^{min,unst}$, whatever the kinetic properties. For these reasons, nucleation very close to the liquidus concentration is physically penalized and hence hard to obtain in our theoretical framework, at least for polymers. This raises the question why the onset of nucleation could be observed as soon as the solubility limit is reached in drying OPV films. [10][17]

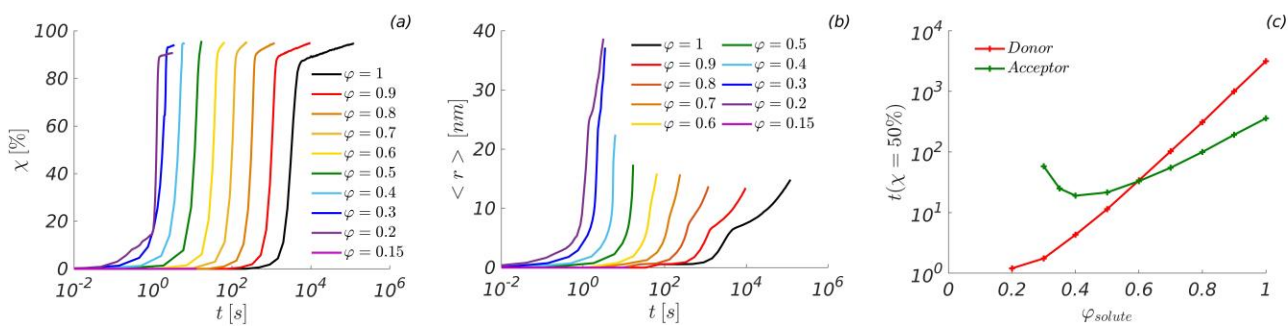

*Figure 2: Crystallization in a binary solute-solvent solution at fixed composition for various solute content. (a) Donor crystallization, time-dependent total amount of crystallized donor. (b) Donor crystallization, time-dependent mean radius of donor crystals. (c) Time required to reach 50% crystallized solute, for both donor and acceptor. The simulation box size is $256\ nm \times 256\ nm$ with a grid spacing of $1\ nm$.*

The average time-dependent radius of the crystals is shown in (**Figure 2b**) for different volume fractions. As expected, the crystals are larger with deceasing super saturation. With the parameters chosen, the crystallization process is significantly dominated by nucleation as compared to classical growth. Nevertheless, the crystals increase due to various phenomena after the build-up phase. In the less dilute solutions, the crystals touch each other and the grain size increases by grain coarsening. In the more dilute solutions, where the crystals are surrounded by the liquid phase, we observe two growth mechanisms, leading to very rapid growth: first, Ostwald ripening causes smaller crystals to disappear to the benefit of the larger ones. Second, the concentration in the liquid phase is decreasing with time, therefore the critical germ size increases. Germs with a radius initially just above the critical radius, and thus initially stable, become unstable and disappear. This means that obtaining small, slowly growing crystals that nucleate early (i.e. at high dilution) is difficult, and that a later nucleation should favour a morphology with smaller crystals.

In **Figure 2c**, the crystallization rates for the donor and the acceptor in a solution at fixed global composition are compared. As can be expected from the phase diagrams (**Figure 1a** and **c**), crystallization sets on at higher concentration for the small-molecule acceptor, but the composition-dependency is less pronounced because of the smaller difference in the diffusion coefficient of pure solvent and pure solute. This results in a faster crystallization process for the pure acceptor as compared to the pure polymer.

In addition to the characterization of the crystallization kinetics, the knowledge of the drying kinetics is required in order to understand the behaviour of the drying film. The expected drying curves, corresponding to the simulations of evaporating films (see next section), but excluding any crystallization process, are shown in **Figure 3**. Thereby, the initial mixture is a $20:13:67$ donor:acceptor:solvent blend (1:1 donor:acceptor weight ratio). The initial height is 450nm and the final height $148\ nm$. Evaporation occurs at a constant rate almost until the end as expected for a polymer solution. The diffusion-limited evaporation phase at the end of the drying is almost absent, because the time scale for diffusion in the liquid remains small as compared to the time scale of the drying: we use relatively high diffusion coefficients of the solvent at high polymer concentration, because at 60°C the systems always remain above the glass transition temperature of P3HT. The corresponding (time-dependent) Biot number $h(t)v_e(t)/D(t)$ is in the range $4.10^{-5}$ to $10^{-1}$ ($h$ being the film height, $v_e$ the surface vertical displacement speed and $D$ the mutual diffusion coefficient). In general, diffusion processes in the liquid are fast as compared to the evaporation and the crystallization processes, so that no concentration gradients due to diffusional limitations are expected. In addition, this means that whenever a spinodal decomposition is possible, the liquid-liquid demixing should be completed almost instantaneously (as compared to the other processes).

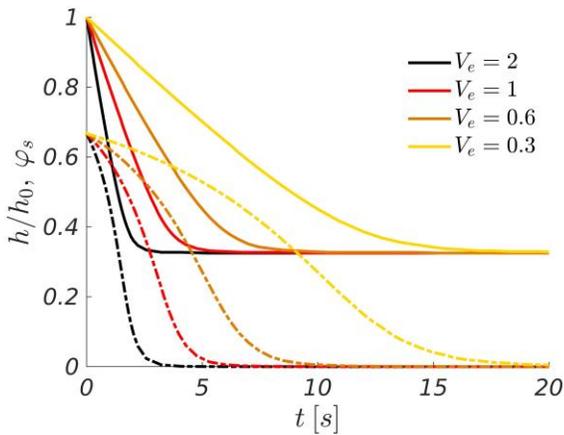

Figure 3: drying curves for the $20:13:67$ donor-acceptor-solvent blend at different evaporation rates such that the evaporation-condensation coefficient is $\alpha = V_e \alpha_{ref}$. (Full lines) normalized film height relative to the initial height $h/h_0$. (Dashed lines) solvent volume fraction $\varphi_s$. 1D simulations without crystallization.

In principle, understanding the morphology formation during drying requires a "convolution" of the drying curves shown in **Figure 3** with the crystallization properties shown in **Figure 2**: once the onset of crystallization is reached, this determines whether the crystallization process can take place and to which extent. For instance, considering **Figure 2a**, if the evaporation time after reaching 20% solute concentration would be less than $0.5\ s$, there would be no time for P3HT crystallization during drying. On the contrary, with long evaporation time, we would expect the crystallization to start and complete as soon as 20% solute concentration is reached. In between, a crystallization process ranging from 20% to higher solute volume fractions would be possible.

With the description of the thermodynamic and kinetic properties given above, the time and length scales can be identified. They are the fundamental elements required to qualitatively understand the morphology formation upon drying, as will be illustrated in the next section. Nevertheless, a drying ternary mixture is a very complicated system. We believe that investigating the crystallization behaviour of binary donor-solvent and acceptor-solvent blends at fixed solvent loading, as described above, is a very useful milestone for the physical understanding of OPV BHJ formation. On the one hand, this is a much simpler system as compared to the drying ternary mixture. First, the complexity due the constantly changing solvent loading is suppressed. Second, additional interactions that could occur in the ternary mixture between the phase transformation phenomena (donor and acceptor crystallization, donor-solvent and acceptor-solvent LLPS, donor-acceptor LLPS) are avoided. On the other hand, as shown above, a lot of crucial information can be obtained from such binary blends, reducing the horizon of possibilities. Therefore, an experimental characterization of the crystallization and/or LLPS kinetics in non-evaporating, binary solutions, if possible, would be highly desirable, not only for the sake of strengthening simulations, but also for the general comprehension of solution-processed OPV.

**Relationship to known material parameters for the P3HT-PCBM-DCB system.** The thermodynamic and kinetic properties detailed above have been obtained with parameters inspired by the P3HT-PCBM-DCB mixture. The melting temperature and the heat of fusion for P3HT are in the range $210 - 240$ °C and $30 - 50\ kJ \cdot kg^{-1}$ (although $15 - 20\ kJ \cdot kg^{-1}$ has also been reported), and for PCBM in the range $280 - 290$ °C and $15 - 20\ kJ \cdot kg^{-1}$, respectively.[73][74][75][76][77][78][79] We use 237 °C, $50\ kJ \cdot kg^{-1}$, 285 °C and $20\ kJ \cdot kg^{-1}$ in this work. The P3HT-PCBM interaction parameter has been measured to be $0.7 - 0.9$ [76][80] from the melting point depression, i.e. at temperatures above 200 °C. Unfortunately, we are not aware of any reliable measurement of interaction parameter values reported for lower temperatures. We use a linear temperature dependence $\chi_{ll,da} = 1.592 - 1.778 \cdot 10^{-3}T$ leading to $\chi_{ll,da} = 1$ at 60°C and $\chi_{ll,da} = 0.6$ at 285 °C. The phase diagram presented in **Figure 1d** has been obtained using additionally $W_d = 75\ kJ \cdot kg^{-1}$, $W_a = 30\ kJ \cdot kg^{-1}$ for the energy barriers and $\chi_{sl,da} = \chi_{ll,da} + 0.3$ although no experimental characterization of these parameters is available. Using the same melting points and heat of fusion values for the P3HT-DCB and PCBM-DCB phase diagrams, the interaction parameters $\chi_{ll,ds}$, $\chi_{sl,ds}$, $\chi_{ll,as}$, $\chi_{sl,as}$ are required. To the best of our knowledge, the only experimental estimates available are $\chi_{ll,ds} \approx \chi_{ll,as} \approx 0.4$ for P3HT and PCBM in DCB.[76] Unfortunately, for P3HT, using this value would lead to far too small solubility values at temperatures above 40°C, and to very high volume fractions for the onset of crystallization $\varphi_d^{nucl,onset}$. Since a decrease of $\chi_{ll}$ leads to higher solubility values and a lower $\varphi_d^{nucl,onset}$, we use $\chi_{ll,ds} = 0.05$. The situation is the other way around for PCBM and we use $\chi_{ll,as} = 77.48/T + 0.34$, leading to $\chi_{ll,as} \approx 0.57$ at 60°C. The interaction parameters in the crystalline regions are set to $\chi_{sl,ds} = 0.4$ and $\chi_{sl,as} = \chi_{ll,as} + 0.1$. Concerning the kinetic properties, the self-diffusion coefficients of each material in each pure material are required. Values for the diffusion coefficients of solvents in themselves are typically around $10^{-9}\ m^2 \cdot s^{-1}$, for polymers in solvents $10^{-10} - 10^{-11}\ m^2 \cdot s^{-1}$,[81] for small molecules in solvents $10^{-9} - 10^{-10}\ m^2 \cdot s^{-1}$. For PCBM in DCB, it has been simulated to be $4 \cdot 10^{-10}\ m^2 \cdot s^{-1}$ at room temperature.[82] The temperature-dependence of the diffusion coefficient of PCBM in P3HT has been investigated, leading to values of about $4 \cdot 10^{-15}\ m^2 \cdot s^{-1}$ at 60 °C.[83] Since at this temperature the system is far above the glass transition temperature, we assume the diffusion coefficient of the solvent in the polymer to be at least of the order of $10^{-14}\ m^2 \cdot s^{-1}$. [84] [85] [86] We choose the remaining diffusion coefficients assuming that they are in general lower for the polymer than for the fullerene, and for the fullerene than for the solvent. In particular, the self-diffusion coefficient of the polymer in itself is set to $10^{-16}\ m^2 \cdot s^{-1}$. Once all these parameters are fixed, the time scale for crystallization has to be fixed using the value of $M_{k,0}$ for both donor and acceptor. The values are chosen so that P3HT crystallization can occur during drying, and PCBM crystallization at the end of the drying as observed experimentally. Note that the crystallization time is about $3000\ s$ for pure P3HT (close to the values obtained for P3EHT at this temperature [87]) and $350\ s$ for pure PCBM, respectively. Regarding evaporation, the vapour pressure of DCB at $60\ °C$ is set to $2\ kPa$ and the pressure in the environment to zero. The only unknown parameter is the evaporation-condensation coefficient $\alpha$ which we adjust to obtain the evaporation rates as measured in [12] for the constant evaporation rate phase. Finally, the surface tension parameters $\kappa_i$ and $\varepsilon_k$ are adjusted to obtain surface tensions in the range $5 - 50\ mJ \cdot m^{-2}$. The value of $\kappa_i$ has a weak impact on the morphology, while the value of $\varepsilon_k$ strongly influences the balance between nucleation and growth. Unfortunately, we are not aware of any data set allowing precise determination of $\varepsilon_k$. $\varepsilon_v$ is chosen so that the LV interface is sufficiently broad to assure numerical stability. However, this leads to overestimated SV surface tensions, which luckily has no impact on the evaporation kinetics.[57] The full parameter set can be found in Supporting Information S1.

The previous paragraph demonstrates that identifying all the required input parameters is a delicate task. Many parameters are difficult to measure or can be measured only indirectly, whereby the equations used for extracting them from the measurements should in principle be self-consistent with the framework used for the simulations. Some parameters cannot be easily measured in the desired temperature or composition range. For others, possible evaluation methods still need to be identified. Finally, even if all parameters were perfectly known, there is no warranty that all observables can be properly recovered using the (in this work relatively simple) free energy and composition dependence of the kinetic properties. Clearly, improving the coherence between experimentally measured data and the material properties used in the simulation is an important research topic for the future. Within the framework used for this paper, there is an overall compromise between using parameters close to the experimental values, having correct solubility values, and getting early (close to the solubility value) nucleating, small and slowly growing crystals.

## Simulations of film drying

**Simulation setup.** In this section, the BHJ formation of a donor-acceptor blend in a drying solution is investigated. The parameters used are the same as in the previous section and we aim at mimicking the behaviour of a P3HT-PC$_{61}$BM-DCB mixture. The donor-acceptor volume fraction blend ratio is $60:40$, which corresponds to a $1:1$ mass ratio and the drying temperature is $60\ °C$. This corresponds to the experimental conditions used for the in-situ characterizations performed by Güldal,[12] but the simulation results will also be compared with the information obtained by Schmidt-Hansberg [10] for a $1:0.8$ mass ratio between $15\ °C$ and $40\ °C$ and by Vegso [14] for a $1:0.66$ mass ratio at room temperature. As stated above, the onset of nucleation, and therefore the beginning of the morphology formation is expected for $\varphi_d^{nucl,onset} \approx 0.2$ so that the initial mixture composition is $20:13:67$. This is much later than in the experiment and leads to a shorter evaporation time in the simulation although the evaporation rate is consistent with experiments. After evaporation, the dry film is maintained at $60\ °C$ for some more time so as to simulate a subsequent annealing sequence at this temperature. The simulations are 2D on a $512 \times 256$ mesh with a grid resolution of $1\ nm$. In order to save computational time, the viscosities used in the calculation are unrealistically high (see Supporting Information S1), at least for high solvent loading. This means that the velocities are underestimated and therefore the potential impact of advection on the BHJ formation might be underestimated at the beginning of the drying. In the presented simulations, the main consequences of advective mass transfer are that crystals reaching the film surface are pushed down towards the substrate, and that crystals tend to stick together due to depletion forces between them.[58] All simulation parameters can be found in the Supporting Information S1.

**Time-dependent behaviour, structure analysis, comparison with experimental measurements.** The qualitative film evolution upon drying and further annealing is shown in **Figure 4**. The first P3HT crystals appear after around $1 s$ of drying (**Figure 4a, g**) and carry on appearing upon further drying, generating a phase separation between a P3HT crystalline phase and a liquid phase containing all the PCBM material ($2\ s$, $3\ s$, **Figure 4b-c**, **h-i**). Although crystallization is a driving force from which a P3HT-free liquid phase is expected at equilibrium (see liquidus line in **Figure 1b** and **c**), this liquid phase still contains some polymer because the equilibrium is not yet reached. The composition evolution of the liquid phase with less and less polymer and solvent causes the liquid mixture to become immiscible (see phase diagram in Supporting Information S2.1) and spinodal decomposition takes place between $3\ s$ and $3.25\ s$ (**Figure 4c-d, i-j**).

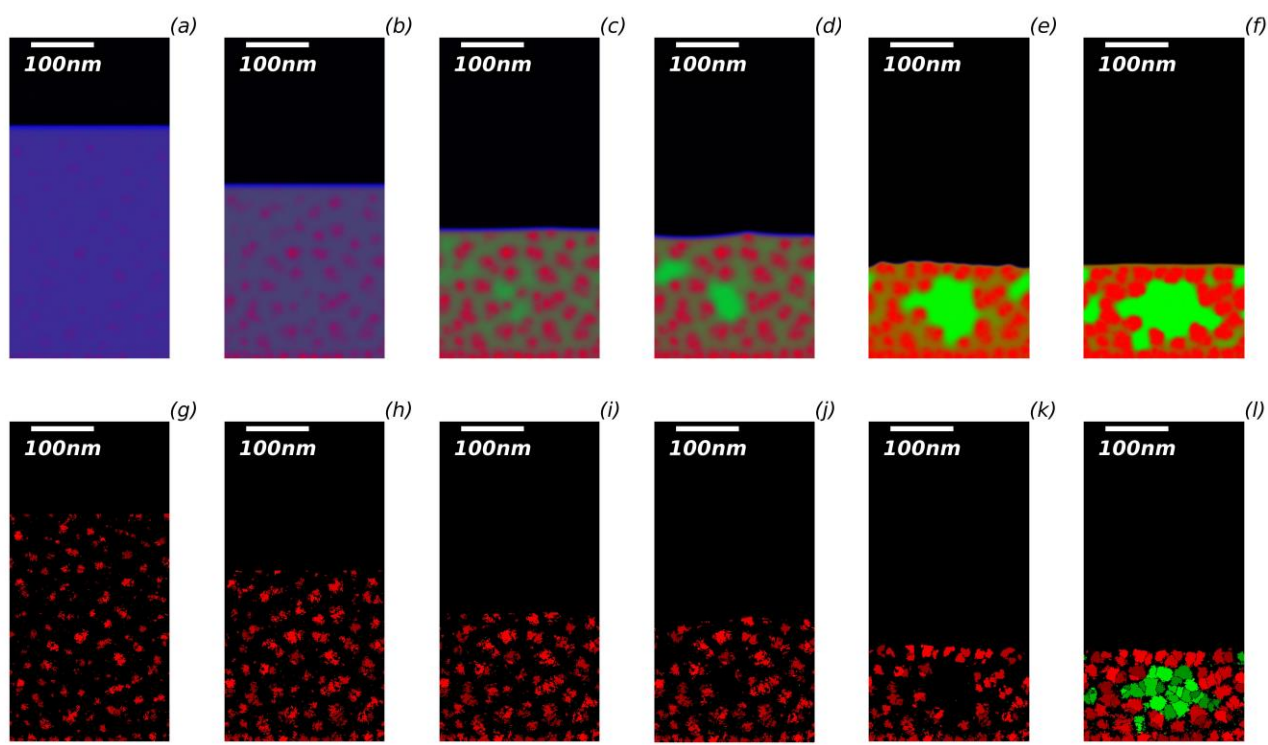

*Figure 4: Drying of a ternary P3HT-PCBM-DCB blend with initial blend ratio is $20 \colon 13 \colon 67$. The volume fraction fields (top row) and the marker fields tracking the presence of the distinct crystallites (bottom row) are shown after 1 s, 2 s, 3 s, 3.25 s, 30 s and 500 s (from left to right). The film is completely dry after 30s. For the volume fraction fields, the polymer donor is represented in red, the acceptor in green and the solvent in blue. For the marker fields, the donor crystals are represented in red and the acceptor crystals in green. The marker value ranges from $-\pi/2$ (dark colour) to $+\pi/2$ (bright colour). The order parameter fields can be found in the Supporting Information S2.3.*

The interfacial energy cost associated with the phase separation is responsible for the fact that pure PCBM phases arise where there is the more free space between crystals. This, together with depletion forces and the minimization of interfacial energy due to concentration gradients between crystals, contributes to the clustering of donor-rich regions and donor crystals into larger-scale, branch-like structures (**Figure 4e-f, k-l**). At the end of the drying, the BHJ is a 3-phase morphology with almost pure P3HT crystallites continuously connected through a mixed amorphous phase, and embedded pure amorphous PCBM areas. The amount of PCBM present in these pure domains is not sufficient to ensure continuity between them. This stresses once again the possible role of the mixed amorphous phase for charge transport.[2] With further annealing, P3HT crystals carry on growing and PCBM crystals appear (**Figure 4f, l**), reducing progressively the amount of amorphous mixed and pure phases. Remember that, since the partial crystallinity of the polymer is not taken into account, the equilibrium morphology is expected to be composed only of pure PCBM and P3HT crystals (see phase diagrams on **Figure 1b** and **d**).

The detailed time-dependent evolution of the film is shown in **Figure 5**. The evolution of the film height during drying is in line with white-light reflectometry (WLR) or laser reflectometry (LR) data.[10][12] The volume fractions of donor, acceptor and solvent in the liquid phase are also shown. The evaporation proceeds at constant rate exactly as shown in **Figure 3** until $3.5s$. At this point, the residual amount of solvent is 15% of the film and further evaporation is hindered by the presence of the crystals, so that the film is fully dry only after 30s. Such a dramatic decrease of the evaporation rate has also been observed experimentally.[12] The time-dependent crystallinity of donor and acceptor materials is also shown in **Figure 5** and can be compared to data obtained by grazing incidence wide angle X-Ray scattering (GIWAXS) [10][12][14] or UV/vis spectroscopy. The morphology formation starts after less than $1\ s$ of drying with P3HT crystallization. The evolution of the solute volume fraction in the liquid is the result of a balance between crystallization (which can be seen as a sink term for the amount of solute in the liquid) and solvent removal (which can be seen as a source term for solute). In the presented case, solvent removal is the dominant mechanism and thus the donor volume fraction in the liquid increases. As a consequence, the mobilities in the liquid phase decrease and the crystallization process becomes slower (see also **Figure 2**) upon drying and almost stops for some period in the dry film. The resulting very progressive time evolution of the crystallinity is remarkably close to the observations made by Güldal, Schmidt-Hansberg and Vegso.[10][12][14]

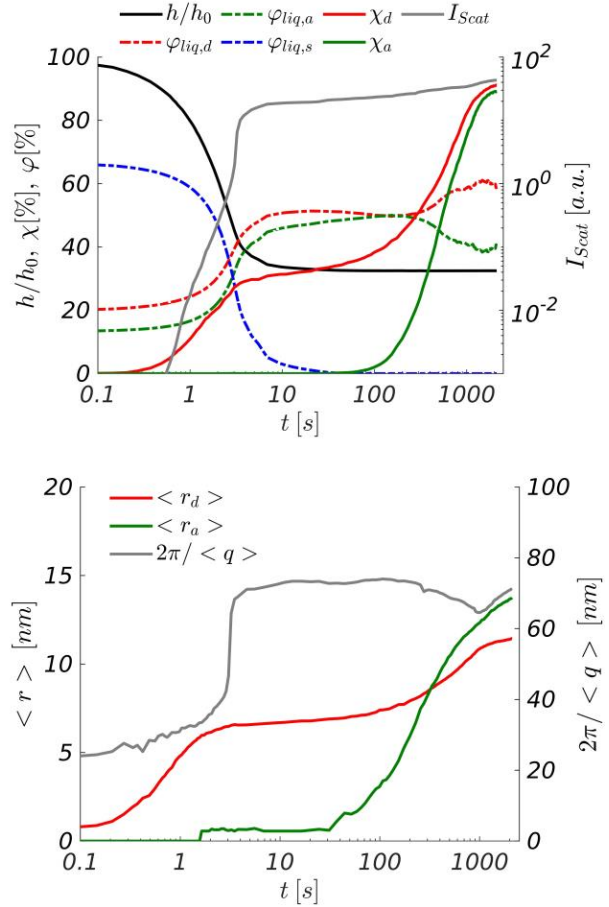

Figure 5: Time-dependent evolution of the donor-acceptor-solvent mixture. (Top) normalized film height relative to the initial height $h/h_0$, donor, acceptor and solvent volume fractions in the liquid phase $\varphi_{liq,d}$, $\varphi_{liq,a}$ and $\varphi_{liq,s}$, percentage of crystalline donor and acceptor materials $\chi_d$ and $\chi_a$ (left y-axis), scattered intensity $I_{Scat}$ for q-vectors corresponding to length scales in the range $[5-150nm]$ (right y-axis). (Bottom) mean radius of donor and acceptor crystals $<r_d>$ and $<r_a>$ (left y-axis), characteristic length scale $2\pi/<q>$ defined from the first moment of the probability distribution of the scattered intensity (right y-axis).

In parallel, the crystal sizes shown in **Figure 5** (bottom) also evolve progressively to reach a radius of $7-8\ nm$ in the dry film, which is comparable to the experimental results obtained from GIWAXS measurements,[12][14] despite of the crude assumption of isotropic growth. The crystals also become more pure due to solvent extraction, as evidenced experimentally by evaluation of the $d_{100}$-spacing (not shown).

The phase separation process is initially driven by the donor crystallization as evidenced by Güldal using light-scattering (LS) measurements on P3HT-PCBM-DCB but also P3HT-DCB mixtures. Although length scales detectable by LS are not accessible in our simulations, scattering data similar to the information contained in grazing incidence wide angle X-Ray scattering diffractograms (GISAXS), for length scales in the range $[5 - 150 \, nm]$, can be obtained. The 2D structure factor of the electronic density field can be calculated at each time step and angular integration gives the wave vector-dependent scattered intensity $I(q)$. Integrating over all $q$-values, or calculating the first moment of this distribution gives the total scattered intensity and a characteristic length scale. Both results are shown in **Figure 5** top and bottom, respectively. As a result of the phase separation, the scattered intensity increases over orders of magnitude as soon as crystallization takes place. Nevertheless, the scattering signal as well as the characteristic length scale abruptly increase with the onset on spinodal decomposition at about $3.2 \, s$. More precise investigation of the diffractograms (see Supporting Information S2.2) reveals a shoulder for $q$-values close to $0.2 \, nm^{-1}$, as observed by Vegso,[14] which documents the phase separation.

Once the film has dried, the structure is kinetically quenched and further evolution requires a significant annealing time. P3HT crystallization proceeds after $100 \, s$ together with PCBM aggregation. After about $2500 \, s$, the film is almost fully crystalline with a mean crystal radius of $11.5 \, nm$ and $13.5 \, nm$ for P3HT and PCBM, respectively. Further changes in the morphology would be due to grain coarsening but would require much longer times (not simulated). The overall composition of the BHJ at the end of the drying and after $500 \, s$ annealing has been analysed and the probability distribution of the donor in the different phases in shown in **Figure 6**.

At the end of the drying, the film is composed of 3 phases: a crystalline P3HT phase representing 23% of the volume, an almost pure PCBM amorphous phase representing 13% of the volume and a majority amorphous mixed phase representing 64% of the volume. The mean volume fraction of P3HT is 0.88 in the crystals, 0.025 in the PCBM amorphous phase and 0.61 in the amorphous mixed phase. This is in line with the current representation of the BHJ in the case of P3HT-PCBM.[2,88] In particular, the amount of amorphous mixed phase and its composition is compatible with charge transport for the electrons and holes. After $500 \, s$, the amounts of crystalline phases are 40% and 19% for PCBM and P3HT, respectively. Due to crystallization, the amount of pure amorphous PCBM phase has dropped to 3% but the amount of mixed phase is still as high as 38%. In particular, the mixed phase is present at interfaces between P3HT and PCBM crystals. The purity of the P3HT crystals increased to a mean volume fraction of 0.92, while the PCBM are significantly purer (0.98 volume fraction of PCBM). The residual pure amorphous phase contains 2% P3HT and the mean composition of the amorphous phase remains unmodified. Further annealing leads to the progressive suppression of the amorphous mixed phase in favour of the crystalline phase with large crystals. Thus, excessive annealing is problematic for the properties of the device, as demonstrated by experiments.[73,77]

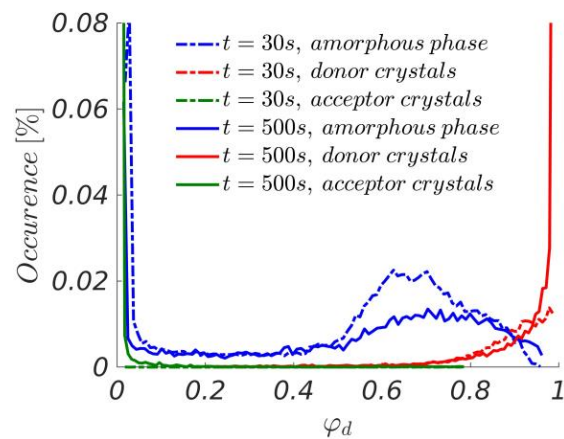

Figure 6: Probability distribution of the donor volume fraction in the amorphous phase, the donor crystals and the acceptor crystals after $30 \, s$ (end of the drying) and after $500 \, s$.

Note that the moment for PCBM crystallization can be a matter of discussion. Indeed, the WAXS signal of PCBM is hidden by the DCB diffusion. Whereas it could not be assessed in the experiments by Güldal, Schmidt-Hansberg found out that PCBM crystallizes shortly before the end of the drying at $25 \, °C$. Such a situation could be easily reproduced in the simulation (see below). However, Vegso suggested that PCBM aggregation occurs before P3HT crystallization (at $25 \, °C$ and for a $1:0.66 \, wt.$ P3HT-PCBM blend), basing on the appearance of $15 \, nm$ structures identified with Guinier analysis of the GISAXS spectra. Unfortunately, there is no direct evidence that this is related to PCBM aggregation. On our side, the low-$q$ range is not accessible and we could not check whether this feature could be recovered in the simulation.

**Summary of the comparison with experiments.** Apart from the semi-crystalline and fibrillary nature of the polymer donor which are not taken into account, we believe that the comparison of the simulation results with experimental data is very promising. As detailed above and summarized in the table below, a lot of different experimental features, obtained with various sophisticated measurement techniques, are reproduced all together in the simulation. We successfully reproduce the morphology formation pathway with P3HT crystallization, subsequent LLPS, and late PCBM crystallization. The time evolution of crystallinity, crystal growth, scattering data, as well as the final morphologies are qualitatively and almost quantitatively compatible with experimental observations. Of course, remember that observations concerning details at the molecular scale, such as the edge-on / face-on orientation of polymer crystals, cannot be taken into account in our continuum mechanics simulations.

| Experimental result | In simulations |
|---|---|
| P3HT crystallization first[10][12] | ✓ |
| Crystallization start @ ~10% vol.[10][12][14] | *~30% wt.* |
| Phase separation due to crystallization[2][12] | ✓ |
| Progressive P3HT crystallization until end of drying[10][12][14] | ✓ |
| Crystal sizes 7-10nm[12][14] | ✓ *(5-7nm)* |
| Increasing crystal purity[12][14] | ✓ |
| PCBM crystallization at end of drying[10] | ✓ |
| Majority mixed amorphous phase[2] | ✓ |
| 20-30%. PCBM in amorphous phase[2] | ✓ *(40%)* |
| Donor/acceptor-rich domain ~50nm[10][14] | ✓ |
| Pathways to electrodes for charge carriers[2] | ✓ |
| Further crystallization upon annealing[10] | ✓ |
| P3HT partial crystallinity[2][88] | **X** |
| P3HT crystals fibrillary structure[88] | **X** |

Table 1: summary of the comparison between experimental and simulation results

**Effect of evaporation rate, acceptor crystallization rate and donor-acceptor miscibility.** The evaporation rate $V_e$, the Allen-Cahn mobility for acceptor crystallization $M_a$ (and thus its crystallization rate), and the Flory-Huggins interaction parameter between donor and acceptor $\chi_{ll,da}$ are varied to investigate their impact on the morphology formation. The corresponding detailed parameter sets can be found in Supporting Information S1. The simulation presented above is used as a reference with the evaporation rate defined as $V_e = 1$, the Allen-Cahn mobility $M_a = 15$ and $\chi_{ll,da} = 1$. The film morphology after $30\,s$ of simulation time is shown in **Figure 7**. The order parameter fields after $30\,s$, the morphology after $500\,s$ as well as the time-dependent film height, liquid phase composition, crystallinity, crystal sizes, scattered intensity and characteristic length scale can be found in the Supporting Information S3.

The impact of the evaporation rate is shown in **Figure 7a-c**, **g-i** as well as **Figure 4e, k** and can be qualitatively understood from the comparison of the crystallization kinetics and the evaporation kinetics (see **Figure 2** and **Figure 3**). Overall, the morphology formation pathway remains the same, starting with the formation of donor crystals followed by phase separation, spinodal decomposition and acceptor crystallization. However, one can observe a significant qualitative change from a progressive, limited crystallization up to the dry state (at high evaporation rates) to an abrupt, full crystallization in a film that does not fully dry (at low evaporation rates). With increasing evaporation rate, there is indeed more and more time available for donor crystallization. Thus, the donor crystallinity and the crystal sizes increase. The balance between the sink and source terms for the donor volume fraction in the liquid phase evolves such that the sink term (crystallization) becomes dominant with lower evaporation rate. For $V_e = 0.3$, the donor volume fraction in the liquid phase only increases from 0.2 to 0.215 and then drops due to crystallization (see Supporting Information S3.3). This results in higher mobilities and hence accelerated crystallization, so that all donor materials fully crystallize and the donor volume fraction in the liquid quickly stabilizes towards the liquid composition. In this case, the crystallization proceeds all of a sudden and crystal sizes also quickly increase. An additional consequence is the accumulation of donor crystals at the surface, fully preventing further evaporation. The film height remains constant, and the high solvent loading ensures high mobilities in the mixture, enabling acceptor crystallization and further donor crystal growth. For $V_e = 0.6$, the evaporation rate and the donor crystallization rate are almost equal. The crystallization process remains progressive and the film can quickly dry until 20% of solvent is remaining. Then, evaporation is strongly hindered and the film is not yet fully dry after $900\,s$. On the other hand, for $V_e = 2$, crystallization is not fast enough and the donor crystallinity at the end of the drying is below 10%.

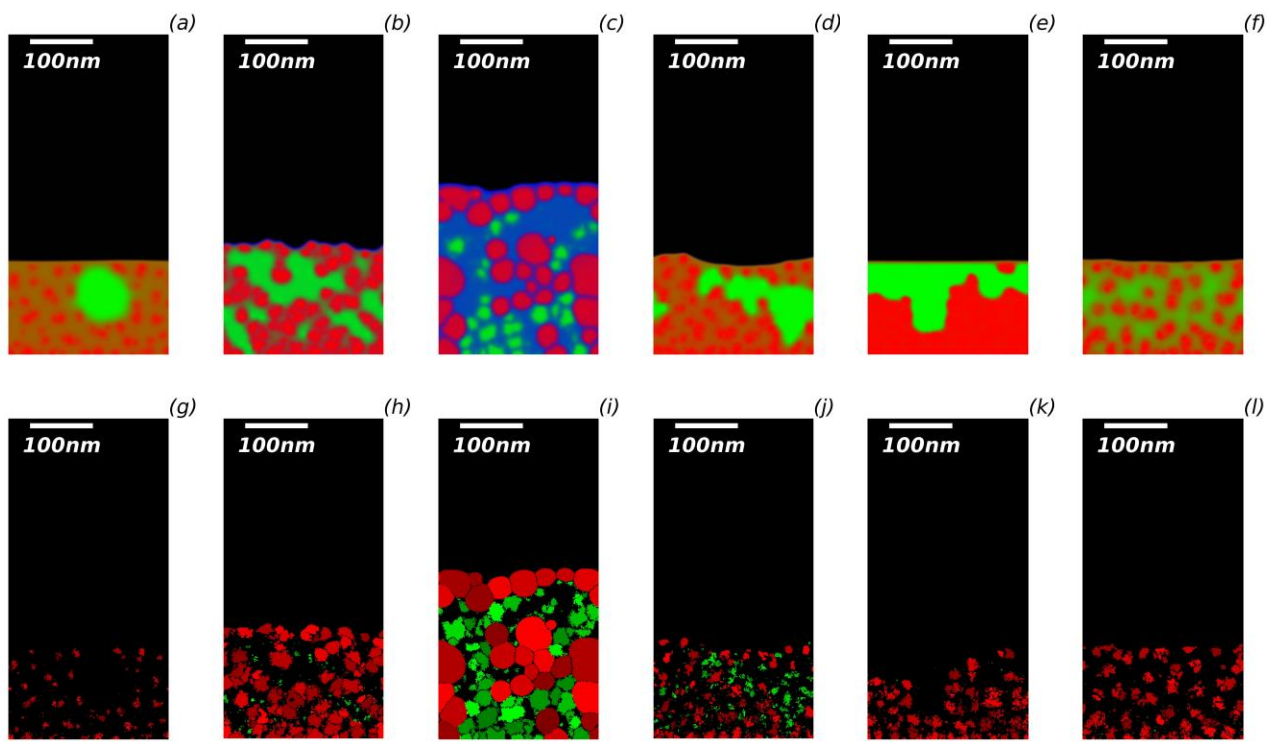

*Figure 7: Morphology of a ternary P3HT-PCBM-DCB blend with initial blend ratio is $20:13:67$ after $30\ s$ drying at $60\ °C$ (except for (c, i): after 16s drying). The volume fractions fields are shown on the top row and the marker fields tracking the presence of the distinct crystallites on the bottom row. From left to right: (a, g) $V_e = 2$, (b, h) $V_e = 0.6$, (c, i) $V_e = 0.3$, (d, j) $M_a = 45$, (e, k) $\chi_{ll,da} = 2.9$, (f, l) $\chi_{ll,da} = 0.6$. For $V_e = 0.6$ and $V_e = 0.3$, the film is not dry. For the volume fraction fields, the polymer donor is represented in red, the acceptor in green and the solvent in blue. For the marker fields, the donor crystals are represented in red and the acceptor crystals in green. The marker value ranges from $-\pi/2$ (dark colour) to $+\pi/2$ (bright colour).*

This means that, with the parameters used in this work, the processing window to obtain a dry film together with donor crystallization during drying is typically one decade of evaporation time. We believe that this is unrealistically small, and we are not aware that such a very sudden crystallization and associated stop of evaporation has been reported for usual OPV systems. This means that the simulations probably have to be improved regarding this topic and this will be considered in future work. Since the switch between progressive and sudden crystallization is the result of a subtle balance between all kinetic and thermodynamic properties, an incorrect prediction can have various reasons, ranging from the assumption of full crystallinity of the polymer donor to a bad estimation of the composition-dependence of the crystallization rate, or an excessive Ostwald ripening rate. A lower evaporation rate also promotes the LLPS during drying. First, as stated before, with slower evaporation for a given film overall composition / height, there is less donor in the liquid phase. Second, this also leads to earlier spinodal decomposition, because the liquid phase is pushed earlier towards the liquid-liquid unstable region of the ternary phase diagram (see Supporting Information S2.1). The consequence of all these phenomena on the final morphology is summarized in **Table 2**. After 30s, the lower the evaporation rate, the higher the crystallinity, the higher the amount of pure amorphous phase, the lower the amount of mixed amorphous phase. For $V_e = 0.6$, the (still wet) film is even fully crystalline after 500s. Therefore, and as can be seen from **Figure 7**, the structure comes close to a co-continuous 2-phase morphology with lower evaporation rate. Otherwise for higher evaporation rates, it can be thought of as a three phase morphology with a majority mixed phase containing donor crystals, separated from pure acceptor islands. It is also worth mentioning that the roughness tends to increase with increasing evaporation time, as experimentally observed by Schmidt-Hansberg, [10] even if more statistics would be required to quantify properly this effect in the simulations.

For dry films, the effect of annealing always obeys the following trends: the crystallinity and crystal sizes increase for both donor and acceptor while the amount of pure and mixed amorphous phase decreases. Since the solidus concentrations are not reached yet, the purity of both crystal phases increase. The composition of the mixed amorphous phase is nearly stable. This is also true for the effect of the acceptor's crystallization rate and of the donor-acceptor miscibility. The main exception is that, with higher acceptor crystallization rate ($M_a = 45$), the donor content in the mixed phase increases because the acceptor crystallization is faster than the donor crystallization.

|  | $V_e = 2$ | | Ref. | | $V_e = 0.6$ | | $V_e = 0.3$ | | $M_a = 45$ | | $\chi_{ll,da} = 2.9$ | | $\chi_{ll,da} = 0.6$ | |
|---|---|---|---|---|---|---|---|---|---|---|---|---|---|---|
| Time [s] | 30 | 500 | 30 | 500 | 30* | 500* | 16* | | 30 | 500 | 30 | 500 | 30 | 500 |
| % donor crystals | 7 | 19 | 23 | 40 | 43 | 61 | 48 | | 17 | 30 | 21 | 35 | 24 | 41 |
| % acceptor crystals | 0 | 11 | 0 | 19 | 1 | 39 | 25 | | 7 | 36 | 0 | 25 | 0 | 11 |
| % pure amorphous | 9 | 3 | 13 | 3 | 16 | 0 | 11 | | 10 | 1 | 32 | 6 | 0 | 4 |
| % mixed amorphous | 84 | 67 | 64 | 38 | 40 | 0 | 16 | | 66 | 33 | 47 | 34 | 76 | 44 |
| $\varphi_d$, donor crystals | 0.86 | 0.92 | 0.88 | 0.92 | 0.79 | 0.84 | 0.66 | | 0.89 | 0.93 | 0.96 | 0.96 | 0.82 | 0.9 |
| $\varphi_d$, acceptor crystals | NA | 0.02 | NA | 0.02 | 0.1 | 0.08 | 0.01 | | 0.22 | 0.07 | 0.06 | 0.01 | NA | 0.04 |
| $\varphi_d$, pure amorphous | 0.03 | 0.02 | 0.03 | 0.02 | 0.02 | NA | 0.01 | | 0.02 | 0.02 | 0.01 | 0.01 | NA | 0.03 |
| $\varphi_d$, mixed amorphous | 0.64 | 0.64 | 0.61 | 0.61 | 0.43 | NA | 0.16 | | 0.67 | 0.88 | 0.84 | 0.85 | 0.53 | 0.53 |

*Table 2: summary of the composition of the BHJ morphology after $30\ s$ and $500\ s$ for different parameter sets. Morphologies marked with a star are not fully dry.*

The most important impact of a faster acceptor crystallization ($M_a = 45$) is that PCBM crystals may appear before the end of the drying (**Figure 7d**, **j**), which has been observed in experiments. [10] In this example, acceptor crystallization remains very limited during drying, but it also partly hinders the donor's crystallization (see **Table 2**, values after $30\ s$). The mixed amorphous phase contains more donor material due to consumption of the acceptor for crystallization. This is particularly clear after annealing ($500\ s$), where there is more crystalline acceptor than crystalline donor available, due to the faster acceptor crystallization rate without solvent.

Finally, the morphologies for a highly incompatible donor-acceptor pair ($\chi_{ll,da} = 2.9$) and an almost compatible one ($\chi_{ll,da} = 0.6$) are shown in **Figure 7e**, **k** and **Figure 7f**, **l**, respectively. For the highest Flory-Huggins interaction parameter, the spinodal decomposition now takes place before the onset on donor crystallization. Acceptor rich phases appear in the upper part (but not at the surface) of the film. Their size is initially about $40\ nm$ radius and very quickly grows to form an acceptor rich layer on top of the film, which leads to the layered structure seen in **Figure 7**. We believe that it is reasonable to think that phases arising from LLPS generate a length scale much longer than the crystals, and that they reach over $100\ nm$ or even micrometres, which is quite common for polymer solutions. Therefore, it might result in a larger scale lateral structuring and could change the result on vertical structuring. Unfortunately, this cannot be observed with our current simulation box size. For the lowest Flory-Huggins interaction parameter, there is no spinodal decomposition during drying or annealing. The only driving force for phase separation is the crystallization process. At the end of the drying, we obtain a 2-phase system with donor crystallites quite homogeneously dispersed in a mixed amorphous phase. The acceptor crystallites and percolated donor crystal network appear upon annealing. As a general trend, spinodal decomposition helps phase separation and hence contributes to the generation of more pure phases, a smaller amount of mixed amorphous phase and a less well-balanced composition of this amorphous phase.

In general, the few examples detailed above show that the film morphology is the result of a complex interplay between the different possible phase transformations, namely evaporation, crystallization of donor and acceptor, and liquid-liquid demixing. This cannot be understood without taking into account the thermodynamic as well as kinetic behaviour of the mixture. As already discussed in the past,[89] it turns out that the equilibrium morphologies are not optimal for the device properties, and that the film structure has to be kinetically quenched in an out-of-equilibrium state. This state should allow not only for efficient charge generation in donor and acceptor phases and fast charge transport in pure crystal areas, but should also provide percolated pathways to the electrodes for the charge carriers.

## Conclusions and perspectives

In this work, we showed how our recently developed coupled fluid mechanics - phase field framework can be used to investigate the formation of bulk heterojunctions upon drying of organic photoactive layers, which is a complex coupled thermodynamic and kinetic problem. The structure formation is driven by three main thermodynamic processes (solvent evaporation, crystallization, liquid-liquid phase separation) and the kinetic evolution of the structure during drying is simulated taking into account the composition dependent kinetic properties of the mixture (diffusion coefficients, viscosity, crystallization and evaporation kinetics). Advection and diffusion are the active mass transfer processes, enabling the structure formation until the structure is kinetically quenched with decreasing solvent content. Such a tool helps to understand why and how a given morphology forms, depending on the complex interplay between thermodynamic and kinetic properties. Many different scenarios are possible for BHJ formation in organic photovoltaics systems and can be simulated. The simulation gives fast and deep insights into the morphology and simple quantitative analysis tools can be applied for comparison with experimental data obtained with various techniques. The qualitative comparison with measurements published in previous works on the well-known P3HT-PCBM system processed with DCB is very promising. Such a simulation framework could be hopefully a very useful tool in the future to increase the control over the process-structure relationships for researchers working in the field.

We do not claim having fully unravelled the BHJ morphology formation mechanisms in general or even for this particular material system. It has been shown that the identification of proper input parameters is complicated and that the match with experimental results is not perfect yet. Some physical trade-offs identified in this work questions the picture of small, early forming crystals (starting from the solubility limit), growing slowly over a large period of the drying process. Moreover, a perfect quantitative match of such simulations with experiments is probably out of reach, due to limitations inherent to the simulation method (continuum mechanics approach, unrealistically thick interfaces or viscosity values due to numerical limitations). Despite of these, the experimentally observed BHJ formation mechanism and interactions between crystallization and LLPS, the time-dependent crystallinity and crystal sizes, as well as the final three phase morphology with a majority mixed amorphous phase featuring pathways to the electrodes for both charge carriers, have been successfully simulated for the first time. The simulated effect of thermal annealing is also in line with experimental observations. This indicates that the main physical phenomena responsible for the BHJ morphology are already caught in the model. Beyond this, several morphology formation pathways are conceivable and can be sorted out with further investigations. The parameter space is huge and many other formation mechanisms are possible, for sure.

We believe that the quality of the simulations can still be improved in the future. On the one hand, the local free energy functional used for the thermodynamic properties and the functional dependencies of kinetic properties on composition are probably the simplest that can be imagined. Taking into account the much more complex behaviour or real material systems is a major research topic for the future. On the other hand, several physical phenomena of crucial importance, especially for such OPV materials, are currently missing in the model and will be implemented in near future, namely the description of material-specific interactions with the substrate, crystal growth anisotropy and semi-crystallinity.

Nevertheless, the priority research topic in near future will be to extend the comparison of simulations and experiments on other material systems, including more recent polymer-small molecule and all-small molecules organic bulk heterojunctions with non-fullerene acceptors (NFA) but also solution-processed perovskite films. To ensure a better understanding of the crystallization behaviour and a better identification of input parameters, measuring, if possible, the crystallization process of each single material in a solvent at fixed concentrations (similar to what has been simulated in **Figure 2**) would be certainly very helpful. In general, for a detailed understanding of the morphology formation, careful measurements of the thermodynamic and kinetic parameters have to be performed, as well as an exhaustive in-situ experimental structure assessment during drying. The morphology descriptors can easily be adapted to various measured quantities. Note that the framework is in principle versatile regarding the processing route and can be used for solvent or thermal annealing steps, or even for investigations on morphological stability during device usage. Systematic studies concerning the impact of process parameters on the morphology formation pathways and the final structure can also be performed. The overarching goal would be to gain control over the process-structure relationship and propose physics-based design rules for the fabrication process. Finally, following previous work,[43,44,90,91,92,93] morphology descriptors related to the optoelectronic performance can be defined for estimation of the structure-property relationships. Applying them to simulated morphologies obtained with various process parameters opens the way to an improved understanding of the impact of the process on the performance of solution-processed solar cells. Therefore, we hope that the results presented in this paper represent the starting point of a fruitful contribution to the development of solution-processed solar cells.

# Conflicts of interest

The authors declare no conflict of interest.

# Acknowledgements

The authors acknowledge financial support by the German Research Foundation (DFG, project HA 4382/14-1) the European commission (project 101008701), and the Impulse and Networking Fund of the Helmholtz Society. They gratefully thank Prof. Dr. Christoph Brabec, Dr. Ning Li, Dr. Hans-Joachim Egelhaaf and Prof. Dr. Tobias Unruh for fruitful discussions as well as Stefan Langner for solubility and solvent data.

# Notes and references

# Supporting Information

# "Formation of crystalline bulk heterojunctions in organic solar cells: insights from phase-field simulations"

Olivier J.J. Ronsin [a,b]* and Jens Harting [a,b,c]

[a] *Forschungszentrum Jülich GmbH, Helmholtz Institute Erlangen-Nürnberg for Renewable Energy (IEK-11), Fürther Straße 248, 90429 Nürnberg, Germany*
[b] *Department of Chemical and Biological Engineering, Friedrich-Alexander-Universität Erlangen-Nürnberg, Fürther Straße 248, 90429 Nürnberg, Germany*
[cf] *Department of Physics, Friedrich-Alexander-Universität Erlangen-Nürnberg, Fürther Straße 248, 90429 Nürnberg, Germany*

**Corresponding Authors**

E-mail: o.ronsin@fz-juelich.de (O.R.)

# S1. Simulation parameters

Unless otherwise specified, the parameters below are used for all simulations of the paper. For the phase diagrams, only the thermodynamic parameters are used. For the simulations of crystallization in a binary mixture at fixed concentration, the parameters relative to evaporation and the gas buffer material are not used. Subscripts 'd', 'a', 's', 'g' stand for 'donor', 'acceptor', 'solvent', and buffer 'gas', respectively.

The parameter highlighted in <span style="color:red">red</span> are the ones that have been varied for the simulations of film drying (see Table S2 below).

| Parameters | Full name | Values | Units |
|---|---|---|---|
| GENERAL | | | |
| $dx, dy$ | Grid spacing | 1 | nm |
| $T$ | Temperature | 333 | K |
| $\rho_d, \rho_a, \rho_s, \rho_g$ | Density | 1100, 1600, 1300, 1300 | kg·m$^{-3}$ |
| $m_d, m_a, m_s, m_g$ | Molar mass | 30, 0.91, 0.147, 0.03 | kg·mol$^{-1}$ |
| $v_0$ | Molar volume of the Flory-Huggins lattice site | 2.3·10$^{-5}$ | m$^3$·mol$^{-1}$ |
| $N_d, N_a, N_s, N_g$ | Molar volume (given as number of lattice site) | 1182, 24.6, 4.9, 1 | - |
| THERMODYNAMICS | | | |
| $\chi_{da,ll}, \chi_{ds,ll}, \chi_{dg,ll}$ $\chi_{as,ll}, \chi_{ag,ll}, \chi_{sg,ll}$ | Liquid-liquid interaction parameter | <span style="color:red">$1.592 - 1.778 \cdot 10^{-3} T$</span>, 0.05, 0 $77.48/T + 0.34$, 0, 0 | - |
| $\chi_{da,sl}, \chi_{ds,sl}, \chi_{dg,sl}$ $\chi_{ad,sl}, \chi_{as,sl}, \chi_{ag,sl}$ | Solid-liquid interaction parameter (1$^{st}$ material in solid state, 2$^{nd}$ material in liquid state) | $\chi_{da,ll} + 0.3$, $\chi_{ds,ll} + 0.35$, 0 $\chi_{da,ll} + 0.3$, $\chi_{ds,ll} + 0.1$, 0 | - |
| $T_{m,d}, T_{m,a}$ | Melting temperature | 510, 558 | K |
| $L_d^{fus}, L_a^{fus}$ | Heat of fusion | 5·10$^4$, 2·10$^4$ | J·kg$^{-1}$ |
| $W_d, W_a$ | Energy barrier upon crystallization | 7.5·10$^4$, 3·10$^4$ | J·kg$^{-1}$ |
| $P_0$ | Reference pressure | 10$^5$ | Pa |
| $P_{sat,d}, P_{sat,a}, P_{sat,s}, P_{sat,g}$ | Vapor pressure | 1.7·10$^2$, 10$^2$, 2·10$^3$, 10$^8$ | Pa |
| $P_d^\infty, P_a^\infty, P_s^\infty, P_g^\infty$ | Partial pressure in environment | 0, 0, 0, 0 | Pa |
| $E_{p,0}, E_{sm,0}$ | Solid-vapor interaction energy | 5·10$^9$, 5·10$^9$ | J·m$^{-3}$ |
| $\beta$ | Numerical free energy coefficient | 10$^{-5}$ | J·m$^{-3}$ |
| $\kappa_d, \kappa_a, \kappa_s, \kappa_g$ | Surface tension parameter for composition gradients | 10$^{-10}$, 10$^{-10}$, 10$^{-10}$, 2·10$^{-9}$ | J·m$^{-1}$ |
| $\varepsilon_d, \varepsilon_a$ | Surface tension parameter for liquid-solid order parameter gradients | 10$^{-5}$, 10$^{-5}$ | (J·m$^{-1}$)$^{0.5}$ |
| $\varepsilon_{g,d}, \varepsilon_{g,a}$ | Surface tension parameter for grain boundaries | 3·10$^{-2}$, 3·10$^{-2}$ | J·m$^{-2}$ |
| $\varepsilon_v$ | Surface tension parameter for liquid-vapor order parameter gradients | 1.5·10$^{-4}$ | (J·m$^{-1}$)$^{0.5}$ |
| KINETICS | | | |
| $D_{s,d}^{\varphi_d \to 1}, D_{s,d}^{\varphi_a \to 1}, D_{s,d}^{\varphi_s \to 1}, D_{s,d}^{\varphi_g \to 1}$ $D_{s,a}^{\varphi_d \to 1}, D_{s,a}^{\varphi_a \to 1}, D_{s,a}^{\varphi_s \to 1}, D_{s,a}^{\varphi_g \to 1}$ $D_{s,s}^{\varphi_d \to 1}, D_{s,s}^{\varphi_a \to 1}, D_{s,s}^{\varphi_s \to 1}, D_{s,s}^{\varphi_g \to 1}$ $D_{s,g}^{\varphi_d \to 1}, D_{s,g}^{\varphi_a \to 1}, D_{s,g}^{\varphi_s \to 1}, D_{s,g}^{\varphi_g \to 1}$ | Self-diffusion coefficients in pure materials (ex: $D_{s,d}^{\varphi_a \to 1}$ is the self-diffusion of the donor in a pure acceptor matrix) | 10$^{-16}$, 5·10$^{-16}$, 5·10$^{-11}$, 5·10$^{-11}$ 4·10$^{-15}$, 10$^{-13}$, 5·10$^{-10}$, 5·10$^{-10}$ 10$^{-14}$, 10$^{-12}$, 2·10$^{-9}$, 2·10$^{-9}$ 10$^{-14}$, 10$^{-12}$, 2·10$^{-9}$, 2·10$^{-9}$ | m$^2$·s$^{-1}$ |
| $M_{d,0}, M_{a,0}$ | Allen Cahn mobility coefficients for crystals | 1.2·10$^{-6}$, <span style="color:red">$M_a$</span>·6·10$^{-4}$ | s$^{-1}$ |
| $M_v$ | Allen Cahn mobility coefficients for the vapor | 10$^6$ | s$^{-1}$ |
| $\alpha$ | Evaporation-condensation coefficient | <span style="color:red">$-V_e$</span>·2.3·10$^{-5}$ | - |
| FLUID MECHANICS | | | |
| $\eta_d, \eta_a, \eta_s, \eta_g$ | Viscosity | $(10^8, 10^5, 10^1, 10^{-1})/V_e$ | Pa·s$^{-1}$ |

*Table S3: thermodynamic and kinetic parameter set used for the simulations presented in the paper.*

Concerning the simulations of film drying, only the parameters $\chi_{da,ll}$, $M_a$ and $V_e$ have been varied, taking following values:

|  | Ref. | $V_e = 2$ | $V_e = 0.6$ | $V_e = 0.3$ | $M_a = 45$ | $\chi_{da,ll} = 2.9$ | $\chi_{da,ll} = 0.6$ |
|---|---|---|---|---|---|---|---|
| $\chi_{da,ll}(333K)$ | 1 | 1 | 1 | 1 | 1 | **2.9** | **0.6** |
| $M_a$ | 15 | 15 | 15 | 15 | **45** | 15 | 15 |
| $V_e$ | 1 | **2** | **0.6** | **0.3** | 1 | 1 | 1 |

*Table S4: parameters that have been varied for the simulations of film drying. For $\chi_{da,ll}$, only the value at 333K (the temperature used for the simulations) is given.*

The model requires some additional parameters for the noise terms, the diffusion coefficients in the vapor phase, detection threshold for the presence of the crystals and the gas phase that are given in the table below. A penalty for the kinetic properties (diffusion coefficients, order parameter fluctuations, viscosity, order parameter mobility in the SV interface) inside the crystals is applied using the penalty function *f* defined by

$$log(f(x,d,c,w)) = \frac{1}{2} log(d) \left(1 + tanh(w(x-c))\right)$$

where *x* is a field variable. The parameter values for the penalties are given in the table below. More details on the meaning of these parameters can be found in [1].

| Parameters | Full name | Values | Units |
|---|---|---|---|
| THERMODYNAMICS | | | |
| $\sigma_{CH}, \sigma_{AC}$ | *Prefactors for noise terms in the Cahn-Hilliard and Allen-Cahn equation* | $10^{-5}$, 1 | - |
| KINETICS | | | |
| $D_d^{vap}, D_a^{vap}, D_s^{vap}, D_g^{vap}$ | *Diffusion coefficients in the vapor phase* | $V_e \cdot (10^{-17}, 10^{-17}, 2 \cdot 10^{-9}, 2 \cdot 10^{-9})$ | $m^2 \cdot s^{-1}$ |
| DETECTION THRESHOLDS | | | |
| $t_{\phi,d}, t_{\phi,a}$ | *Threshold for crystal detection related to order parameter values* | 0.4, 0.4 | - |
| $t_{\varphi,d}, t_{\varphi,a}$ | *Threshold for crystal detection related to volume fraction values* | 0.2, 0.2 | - |
| $t_{\phi v}$ | *Threshold for vapor detection related to order parameter values* | 0.02 | - |
| PENALTY FUNCTIONS | | | |
| $d_{sl}, c_{sl}, w_{sl}$ | *Amplitude, center and width of the penalty function for diffusion coefficients upon solid-liquid transition* | $10^{-6}$, 0.97, 35 | - |
| $d_\zeta, c_\zeta, w_\zeta$ | *Amplitude, center and width of the penalty function for order parameter fluctuations* | $10^{-2}$, 0.85, 15 | - |
| $d_\eta, c_\eta, w_\eta$ | *Amplitude, center and width of the penalty function for the viscosity upon solid-liquid transition* | $10^{-7}$, 0.2, 20 | - |
| $d_{sv}, c_{sv}, w_{sv}$ | *Amplitude, center and width of the penalty function for Allen-Cahn mobility and interaction energy upon solid-vapor transition* | $10^{-3}$, 0.3, 15 | - |

*Table S5: thermodynamic and kinetic parameter set used for the simulations presented in the paper.*

# S2. Supporting information for the reference simulation

This section provides some details regarding the simulation presented in the section 'Time-dependent behavior, structure analysis, comparison with experimental measurements' of the main text.

## S2.1 Ternary phase diagram without acceptor crystalline state

The ternary phase diagram can also be calculated without considering the crystalline state for the acceptor. This is useful for the understanding of the film behavior upon drying, because the acceptor kinetically does not have time to crystallize, so that the film behaves as if the acceptor were amorphous.

The unstable domain on the left with a maximum around 25% donor and 60% acceptor corresponds to the liquid-liquid demixing. Once the composition of the liquid phase reaches this region, it undergoes a spinodal decomposition.

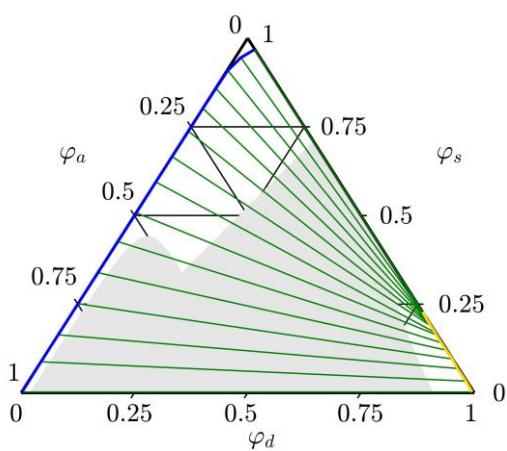

*Figure S8: ternary donor-acceptor-solvent phase diagram at $60\,°C$ of the investigated mixture, not taking into account acceptor crystallization. The unstable compositions are represented in grey, the liquids in blue, the acceptor solids in orange, the donor solids in yellow and the tie lines in green.*

## S2.2 Kratky diffractograms for the reference simulation

The figure below shows the diffractograms of the scattered intensity depending on the wave vector $q$ (after angular integration) in the Kratky representation. During the drying phase, a clear structure appears around $0.2\,nm^{-1}$.

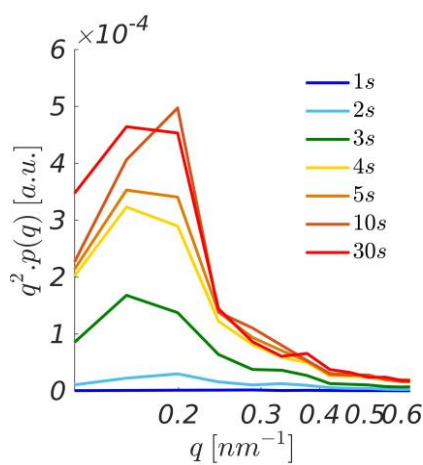

*Figure S9: Kratky plot of the scattered intensity during the drying phase.*

## S2.3 All phase fields for the time-dependent film morphology

The figure below shows the order parameter fields (upper row) for the simulation presented in the section 'Time-dependent behavior, structure analysis, comparison with experimental measurements' of the main text. The volume fraction and marker fields presented in Figure 4 of the main text are reported in the figure below for clarity (top and middle row, respectively).

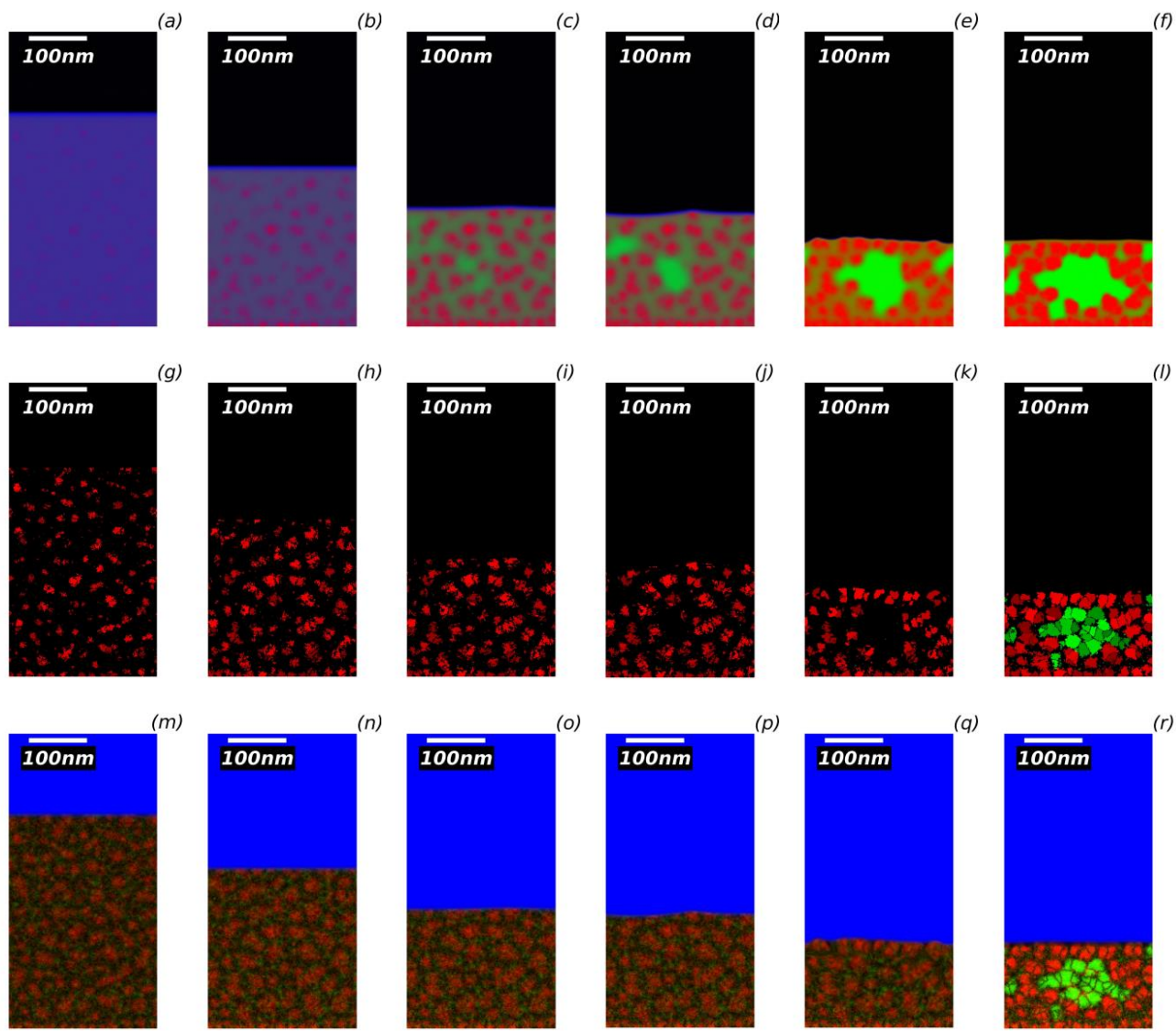

*Figure S10: drying of a ternary P3HT-PCBM-DCB blend with initial blend ratio $20\!:\!13\!:\!67$. The volume fraction fields (top row), the marker fields tracking the presence of the distinct crystallites (middle row) and the order parameter fields (bottom row) are shown after $1\,s$ (a, g, m), $2\,s$ (b, h, n), $3\,s$ (c, i, o), $3.25\,s$ (d, j, p), $30\,s$ (e, k, q) and $500\,s$ (f, l, r) (from left to right). The film is completely dry after $30\,s$. For the volume fraction fields, the polymer donor is represented in red, the acceptor in green and the solvent in blue. For the marker fields, the donor crystals are represented in red and the acceptor crystals in green. The marker value ranges from $-\pi/2$ (dark color) to $\pi/2$ (bright color). For the order parameter fields, the polymer donor crystals are represented in red, the acceptor crystals in green and the vapor phase in blue.*

# S3. Supporting information for the simulations with parameter variations

This section provides further details on the simulation presented in the section 'Effect of evaporation rate, acceptor crystallization rate and donor-acceptor miscibility' of the main text.

## S3.1 All phase fields for the film morphology after 30 $s$

The figure below shows the order parameter fields corresponding to Figure 7 of the main text. The volume fraction and marker fields are reported in the figure below for clarity.

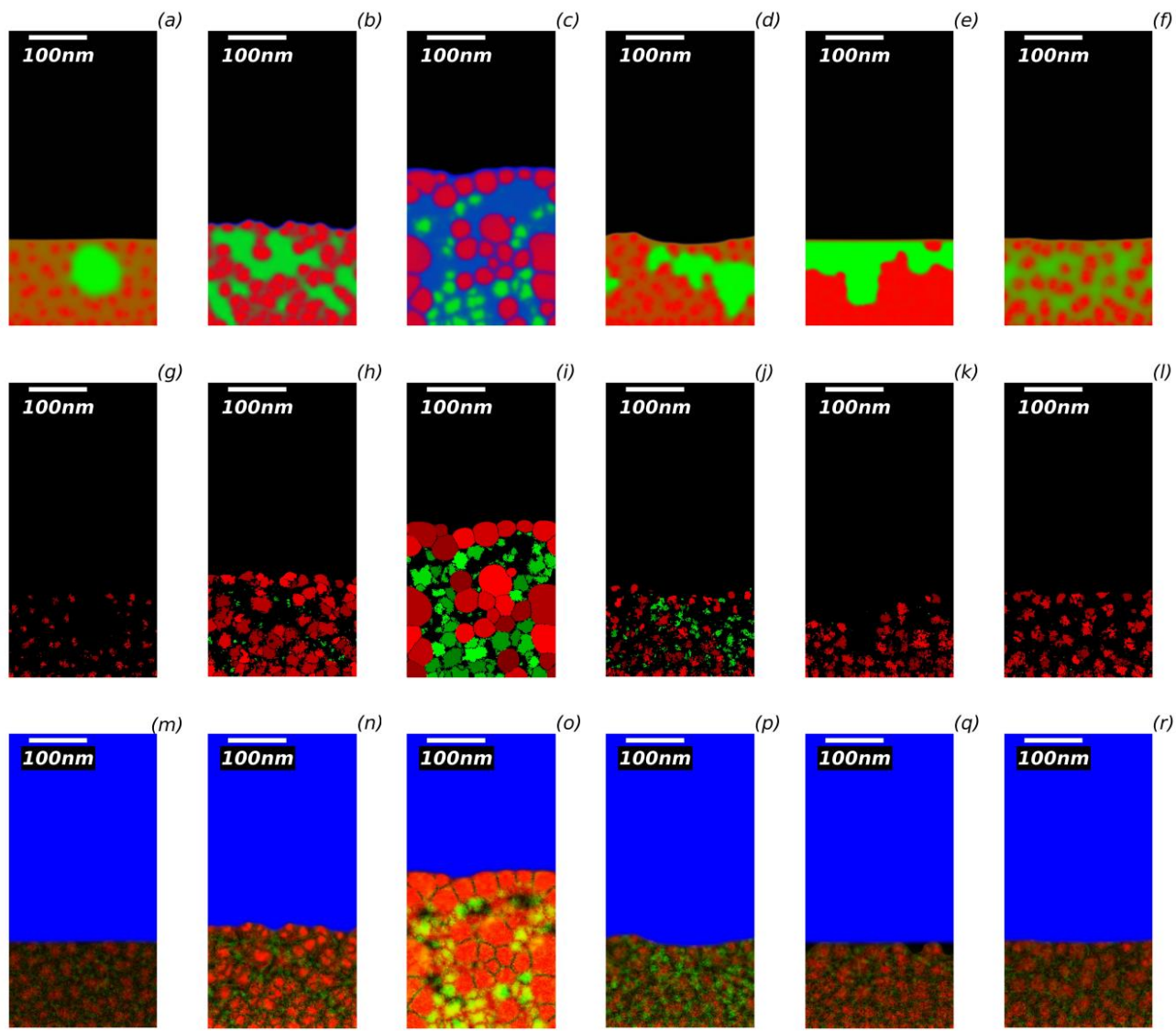

*Figure S11: morphology of a ternary P3HT-PCBM-DCB blend with initial blend ratio is* $20:13:67$ *after* $30\ s$ *drying at* $60\ °C$ *(except for (c, i, o): after 16s drying). The volume fractions fields are shown on the top row, the marker fields tracking the presence of the distinct crystallites on the middle row and the order parameter fields on the bottom row. From left to right: (a, g, m)* $V_e = 2$*, (b, h, n)* $V_e = 0.6$*, (c, i, o)* $V_e = 0.3$*, (d, j, p)* $M_a = 45$*, (e, k, q)* $\chi_{ll,da} = 2.9$*, (f, l, r)* $\chi_{ll,da} = 0.6$*. For* $V_e = 0.6$ *and* $V_e = 0.3$*, the film is not dry. The color code is the same as compared to Figure 3.*

## S3.2 All phase fields for the film morphology after 500 $s$

The figure below shows all phase fields for the simulations presented in the section 'Effect of evaporation rate, acceptor crystallization rate and donor-acceptor miscibility' of the main text, but after 500 $s$. The case $V_e = 0.3$, for which evaporation is blocked by crystals at the film surface, is not represented.

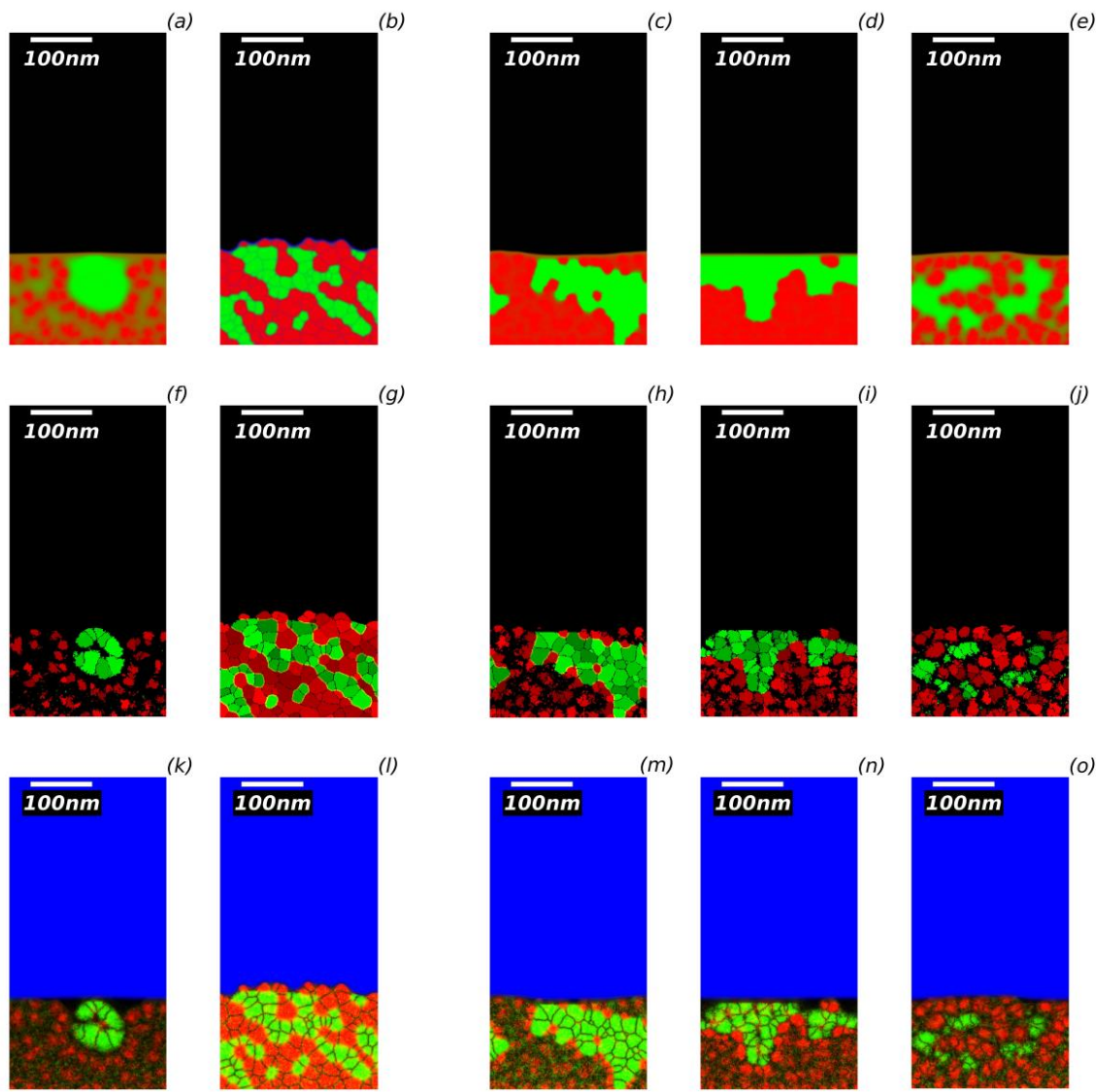

*Figure S12: morphology of a ternary P3HT-PCBM-DCB blend with initial blend ratio is* $20:13:67$ *after* $500\ s$ *drying at* $60\ °C$. *The volume fractions fields are shown on the top row, the marker fields tracking the presence of the distinct crystallites on the middle row and the order parameter fields on the bottom row. From left to right: (a, f, k)* $V_e = 2$, *(b, g, l)* $V_e = 0.6$, *(c, h, m)* $M_a = 45$, *(d, i, n)* $\chi_{ll,da} = 2.9$, *(e, j, o)* $\chi_{ll,da} = 0.6$. *For* $V_e = 0.6$ *the film is not fully dry. The color code is the same as compared to Figure 3.*

s28

## S3.3 Time dependent characterization of the film height, composition and morphology for various parameters

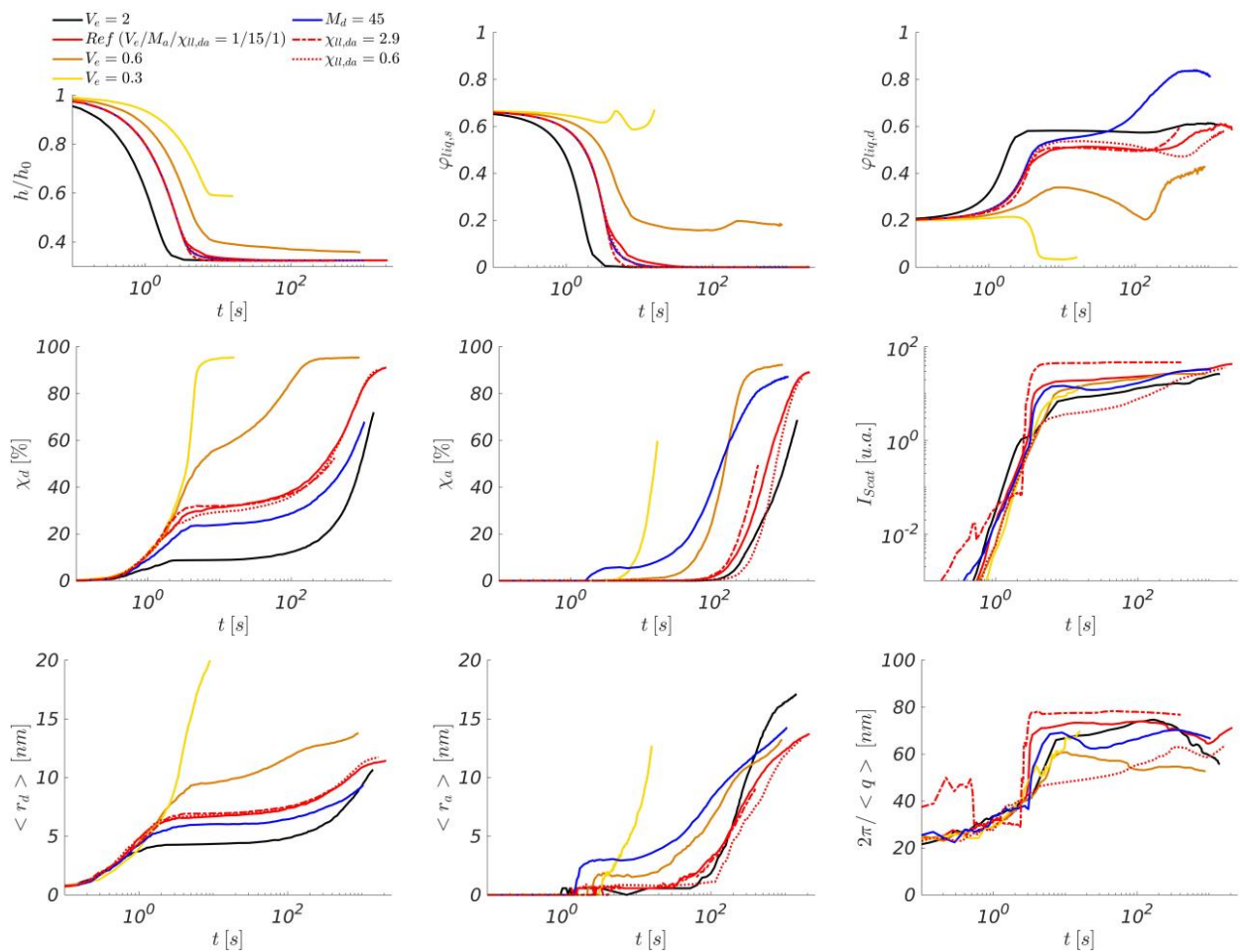

*Figure S13: time-dependent evolution of the donor-acceptor-solvent mixture for various simulation parameters. (Top row, from left to right) Normalized film height relative to the initial height $h/h_0$, solvent and donor volume fractions in the liquid phase $\varphi_{liq,d}$ and $\varphi_{liq,s}$. (Middle row, from left to right) Percentage of crystalline donor and acceptor materials $\chi_d$ and $\chi_a$, scattered intensity $I_{Scat}$ in the range $[5 - 150\ nm]$. (Bottom row, from left to right) Mean radius of donor and acceptor crystals $<r_d>$ and $<r_a>$, characteristic length scale $2\pi/<q>$ defined from the first moment of the probability distribution of the scattered intensity.*